\documentclass[graybox, envcountchap]{svmult}

\usepackage{mathptmx}        
\usepackage{amsmath}
\usepackage{amssymb}
\usepackage{color}
\usepackage{helvet}          
\usepackage{bbding}
\usepackage{pifont}
\usepackage{courier}         
\usepackage{dirtree}

\usepackage{makeidx}        
\usepackage{graphicx}        
\usepackage{subfig}

\usepackage{multicol}        
\usepackage[bottom]{footmisc}

\usepackage[misc]{ifsym}
\usepackage[numbers]{natbib}
\usepackage{hyperref}        
\hypersetup{citecolor=blue,colorlinks=true,urlcolor=blue,linkcolor=blue}
\usepackage{comment}
\usepackage{newtxtext,newtxmath}
\usepackage{ulem}  

\DeclareTextFontCommand{\emph}{\it}

\newcommand{\cmark}{\ding{51}}%
\newcommand{\xmark}{\ding{55}}

\def\lesssim{\mathrel{\hbox{\rlap{\hbox{\lower4pt\hbox{$\sim$}}}\hbox{$<$}}}}
\def\gtrsim{\mathrel{\hbox{\rlap{\hbox{\lower4pt\hbox{$\sim$}}}\hbox{$>$}}}}
\newcommand{\bea}{\begin{eqnarray}}
\newcommand{\eea}{\end{eqnarray}}

\newcommand{\bP}{{\bf P}}
\newcommand{\bF}{{\bf F}}
\newcommand{\bU}{{\bf U}}

\newcommand{\prim}{{{\mathbf{P}}}}

\newcommand{\Omegabin}{\Omega_\mathrm{bin}}

\makeindex             

\begin{document}


\title{Accretion onto supermassive black hole binaries}
\author{Eduardo M. Guti\'errez, Luciano Combi, and Geoffrey Ryan}
\institute{Eduardo M. Guti\'errez (\Letter) \at Institute for Gravitation and the Cosmos, The Pennsylvania State University, University Park PA 16802, USA \\
Department of Physics, The Pennsylvania State University, University Park PA 16802, USA \\
\email{emgutierrez@psu.edu}
\and Luciano Combi \at Perimeter Institute for Theoretical Physics, Waterloo, Ontario, Canada, N2L 2Y5 \\
Department of Physics, University of Guelph, Guelph, Ontario, Canada, N1G 2W1 \\
\email{lcombi@perimeterinstitute.ca }
\and Geoffrey Ryan \at Perimeter Institute for Theoretical Physics, Waterloo, Ontario, Canada, N2L 2Y5 \\
\email{gryan@perimeterinstitute.ca}
}
%
%
\maketitle

\abstract{
    In this chapter, we give an overview of our current understanding of the physics of accreting massive black hole binaries (MBHBs), with a special focus on the latest developments in numerical simulations and General-Relativistic Magnetohydrodynamics (GRMHD) simulations in particular. We give a self-contained global picture of how to model accretion onto MBHBs, analyzing different aspects of the system such as the dynamics of the circumbinary disk, mini-disks, outflows, the role of magnetic fields, and electromagnetic signatures.
    We discuss important questions and open problems related to these systems, what are the advantages and disadvantages of the different numerical approaches, and what robust knowledge we have built from simulations. 
}


\section{Introduction}
\label{sec:intro}

The numerous detections of gravitational wave (GW) signals made in the last few years by the LIGO\footnote{Laser Interferometer Gravitational-Wave Observatory; \url{https://www.ligo.org/}}, Virgo\footnote{\url{https://www.virgo-gw.eu/ }} and KAGRA\footnote{Kamioka Gravitational Wave Detector; \url{https://gwcenter.icrr.u-tokyo.ac.jp/en/}} collaborations \citep{LIGO2016, LIGO2023_GWTC3} have opened a new and unique observational window to the Universe, giving start to a new era of multimessenger astrophysics.
The enormous potential of these observations was demonstrated with the event GW170817 \citep{LIGO2017_GW170817_EM}, where the GW signal detected from the merger of two neutron stars was followed by the detection of multi-wavelength electromagnetic (EM) radiation on timescales that ranged from seconds to years after the merger \citep{LIGO2017_GW170817_EM}.
This sole breakthrough marked an enormous step forward in our understanding of the most violent events in our Universe.

The vast majority ($> 90\%$) of the detected GW signals to date involved only black holes, and none of them was accompanied by the detection of a clear EM counterpart \citep{LIGO2023_GWTC3}.
This is not surprising: black holes only shine provided they are accreting matter from their surroundings \citep{LyndenBell1969}.
In the case of galactic stellar-mass black holes, this usually happens when they are fed by a nearby companion star, either through winds or through the overflow of its Roche lobe.
The merger of two stellar black holes, however, generally occurs at a later evolutionary stage of a binary system, when the material to feed either black hole has already been exhausted.
It is reasonable then to expect mergers of stellar-mass black holes to be ``dry'', that is, without enough matter around them that makes them accrete and shine.
The chance of stellar-mass black hole binaries to be EM bright during the merger may significantly increase if they are located in a particularly dense medium, for example, within the accretion disk of an active galactic nucleus (AGN) \citep{Bartos_etal2017, Stone_etal2017, McKernan_etal2019, Wang_etal2021, Kimura_etal2021, RodriguezRamirez_etal2024}.
    
The situation is much more promising for \emph{mergers of supermassive black holes}.
Massive black hole binaries (MBHBs) were first proposed to exist in the 1980s with the pioneering works by Begelman et al. \citep{Begelman1980} and Roos \citep{Roos1981}, who suggested that as a by-product of the coalescence of two galaxies, the supermassive black holes (SMBHs) at their centers could end up sinking towards the common gravitational potential of the merged galaxy and form a bound pair.
Dynamic processes in the galactic center, mediated by stars or gas, may extract angular momentum from the binary and reduce the orbital separation to $\sim$ milli-parsec scales \citep{Merritt2005}. 
The orbital movement of such massive objects during the inspiral releases enormous amounts of energy as GWs; the merger itself is among the most luminous events in the Universe \citep{BurkeSpolaor_etal2019}.

The prospect of detecting EM counterparts to GWs from MBHBs increases (compared to stellar black holes) mainly due to two reasons.
First, accretion is stronger the larger the central mass: for fixed conditions of the ambient medium (rest-mass density, temperature, etc.), the accretion rate towards a central gravitating object of mass $M$ scales as the area of the Bondi sphere\footnote{The Bondi sphere determines the outer boundary of the region where the gravitational influence of an object dominates the dynamics of matter. It is given by $\sim 4\pi r_{\rm B}^2$, where $r_{\rm B}\approx 0.1 (kT/{\rm keV})^{-1} (M/10^6M_\odot)~{\rm pc}$ is the Bondi radius \citep{Bondi:1952ni}.} $\sim M^2$.
Second, these sources would be located in the center of galaxies that underwent a merger sometime in the past.
Such an environment may present abundant amounts of gas in their nuclei to feed the black hole pair.
Hence, in principle, an MBHB could be surrounded by a bright accretion flow in a similar fashion to normal AGNs, which makes these sources highly promising multimessenger sources of GWs and electromagnetic waves.

Without direct observational evidence, 
advancements in the field have been mainly driven by theoretical studies.
The study of binary black hole accretion adds several difficulties to the already complex field of accretion onto compact objects.
Binary accretion covers a particularly large parameter space, including the accretion rate $\dot{M}$, total mass $M$, disk scale height $h/r$ and equation of state, viscosity $\nu$ or magnetic field $B$, as well as the binary separation $a$, mass ratio $q$, orbital eccentricity $e$, disk inclination $i$, and black hole spins $\vec{a}_1$, $\vec{a}_2$, all of these potentially playing a role in the accretion dynamics and resulting EM emission. 
\emph{The highly nonlinear and dynamical nature of the problem forces us to depend much more on numerical simulations, which in turn are more expensive and difficult to perform than in single black hole accretion}. The large majority of work on binary black hole accretion has relied on 2-D (vertically integrated) hydrodynamical simulations with Newtonian physics (see \citep{duffell2024santa} and references therein).
A much fewer number of works used three-dimensional General Relativistic Magnetohydrodynamical (GRMHD) simulations, and in general they have explored a subrange of the parameter space or are limited in total duration due to the computational cost \citep{Noble12, Gold_2014a, Gold_2014b, Farris:2012PhRvL, Bowen_etal2018, Bowen2019, noble2021, paschalidis2021minidisk, cattorini2021fully, cattorini2022misaligned, combi2022minidisk, avara2023accretion, bright2023minidisc, Ruiz_etal2023}.

Several reviews have come out in recent years overviewing MBHBs, each of them focusing on different aspects of the problem: formation and evolution of MBHBs \citep{Colpi2014, Dotti2012}, Newtonian \citep{Schnittman2013} and relativistic \citep{Gold2019, CattoriniGiacomazzo2024} aspects of the accretion process, circumbinary disk physics \citep{LaiMunoz2023}, and EM counterparts \citep{DeRosa_etal2019, Bogdanovic_etal2022, DOrazioCharisi2023}.
This chapter aims to be complementary to these and tackle a subset of the problems that accretion onto MBHB rises.
First, we discuss the main modeling approaches taken in the literature, for which we present the differences, advantages, and disadvantages among them.
Second, we choose a few specific questions or open problems related to accretion onto MBHBs on the way to the merger.
In particular, we focus on relatively short separations $a<100\,M$, where $a$ is the major semi-axis of the orbit, and where general relativity (GR) becomes important to describe some aspects of the accretion flow.

The remainder of this chapter is structured as follows.
In Section \ref{sec:MBHB_formation}, we give a brief overview of our current understanding of how MBHBs may form and their GW emission, and in Section \ref{sec:accretion_simple}, we summarize the main known phenomenology of accretion onto these systems.
In Section \ref{sec:modeling_MBHBs}, we discuss the various numerical approaches that have been taken in the field to investigate MBHBs, what are the main difficulties in doing so, what are the key differences with the modeling of single black hole accretion disks, and what are the main approximations, advantages, and disadvantages of the different approaches, emphasizing what we can learn from each of them.
Then, in Section \ref{sec:accretion}, we analyze various of the most compelling aspects related to the physics of the circumbinary disk and the mini-disks, and in Section \ref{sec:outflows}, we discuss the potential outflows.
In Sec. \ref{sec:em}, we discuss what are the main EM signatures that may be associated with each of these components.
Finally, in Sec. \ref{sec:conclusions}, we analyze what are the main open questions and what needs to be accomplished in the theoretical modelling of accreting MBHBs to answer them.

\section{Formation of massive black hole binaries and gravitational wave emission}

\label{sec:MBHB_formation}

Massive galaxies form by successive mergers of smaller galaxies in a hierarchical process \citep{Blumenthal_etal1984}. The relationship between the merger of the galaxy and that of the SMBHs they host is however difficult to establish through observations.
Both processes are temporally separated by millions or billions of years and, in turn, the coalescence of SMBHs occurs in an extremely short period compared to the overall evolution of the merged galaxy.

Several aspects of the evolution of SMBH pairs in a merging galaxy are still under debate.
We give here a brief overview of the various stages that such a pair needs to follow before merging and the main physical processes driving the evolution.
For a more in-depth discussion on this, see Ref. \cite{Colpi2014}.

\begin{enumerate}

     \item {\bf Galaxy merger ($t \sim $ a few Gyr):} Two colliding galaxies merge into a single galaxy within a few orbital periods, typically $t_{\rm orb} \sim 2\pi \sqrt{a^3/GM} = 1.5 \times (M/10^{11} M_\odot)^{-1/2} (a/30 ~ \mathrm{kpc})^{3/2} ~\mathrm{Gyr}$.  
     Initially, the bulges of these galaxies, which contain the SMBHs, behave dynamically as individual objects.
      When the bulges of the two galaxies get close enough, they merge and their nuclei begin to dynamically interact with each other.
     At this point, the two black holes orbit separately in the gravitational potential of the merged galaxy.
     \item {\bf Dynamical friction ($t \sim 10^7-10^9 ~ {\rm yr}$):} 
     The interaction of the black holes with the stars and gas exerts an effect of ``dynamical friction'' on them that slows them down and brings them closer together \citep{MilosavljevicMerritt2001}.
     Assuming a spherically symmetric distribution of stars in the nucleus, this process is capable of reducing the distance between the two black holes from a few hundred pc to $\sim$ a pc.
     This occurs over a timescale of $t_{\rm df}\sim 2 \times 10^8 \ln^{-1} N \times (M/10^6M_\odot)^{-1}(a/100~{\rm pc})^2 (\sigma_* / 100 ~{\rm km~s}^{-1})~{\rm yr}$, where $N$ is the number of background stars, and $\sigma_*$ is their dispersion velocity.

     Once the SMBHs are sufficiently close to each other, at a separation $a\sim 0.1 q/(1+q)^2 (M/10^6M_\odot) (\sigma_*/100{\rm km~s}^{-1})^{-2}~{\rm pc}$, the mass of the gas and stars contained in their orbit is less than the black hole masses and dynamical friction becomes inefficient.
     At this stage, the pair forms a \emph{gravitationally bound system} but its orbit can stall and stop shrinking.
     The dilemma of whether the orbit continues shrinking further down to a merger or not is known as the \emph{``last parsec problem''} \citep{Begelman1980, MilosavljevicMerritt2003}
     \item {\bf Evolution in the last parsec ($t\sim$ uncertain):} There is great uncertainty about the dominant orbital evolution mechanism when the separation of the SMBH pair reaches a parsec or less.
     The current understanding of this problem considers that it may be solved by relaxing some of the assumptions in the simple spherical dynamical friction model, such as considering a non-spherical gravitational potential \citep{Berczik_etal2006, Preto_etal2011, Khan11, Vasiliev_etal2015} or rotation of the galactic nucleus \citep{AmaroSeoane_etal2010a, Sesana_etal2011, HolleyBockelman_Khan2015, Mirza_etal2017, RasskazovMerritt2017}.  Alternatively, additional interactions may play a role, such as the presence of more than two SMBHs \citep{HoffmanLoeb2007, Amaro-Seoane_etal2010b, Ryu_etal2018, Bonetti_etal2019} or interaction with disk gas \citep{Lodato09, Cuadra2009, Roedig_etal2011}. 
     \item {\bf Relativistic regime ($t\sim$ variable):} If the black hole pair reaches separations of $\sim 10^{-2}$--$10^{-3}$ pc, the orbital evolution becomes dominated by the emission of GWs, and the merger typically occurs in less than a Hubble time \citep{Milosavljevic2005}.
     
\end{enumerate}
\vspace{\topsep}

     From the point of view of the GW emission, the relativistic regime is divided into three periods: the \emph{inspiral}, where the SMBHs slowly approach in a quasi-stationary way, the \emph{merger} itself, highly nonlinear, and the \emph{ring-down}, where the remnant compact object emits in the form of GWs the energy stored in the asymmetries of quadrupole order or greater while it relaxes to equilibrium.
     In the fraction of time around the merger, the GW luminosity can be so high that it exceeds the EM luminosity of the entire Universe.
The frequency of these GWs can cover a very broad range, from $\sim {\rm nHz}$ frequencies for masses of $\sim 10^{9-10}M_\odot$ during the inspiral, up to $100$ mHz for masses of $\sim 5\times 10^4M_\odot$ during the merger itself.

The lowest frequency GW from massive binaries may be detected using a {\it Pulsar Timing} array (PTA) \citep{Sazhin1978, Detweiler1979}: an array of many milli-second pulsars which are reliable rotators acting as precise clocks in the sky. The passing of a GW through the galaxy can be detected by measuring carefully the time of arrival of each pulse. Pulsar timing array collaborations including the NANOGrav collaboration have recently found evidence of an ultra-low frequency ``stochastic background'' of GWs \citep{NANOGrav2023}.
The leading hypothesis for the origin of such a stochastic background is that it is the result of the collective effect of all the MBHBs in the Universe \citep{NANOGrav2023_SMBHBs, RajagopalRomani1995, Sesana_etal2008, BurkeSpolaor_etal2019}.
If this is the case, this discovery marks the first direct evidence of the presence of MBHBs in the inspiralling regime.
In the next decade, PTAs will continue collecting data, increasing the statistical significance of their results.
This will hopefully enable the detection of the continuous GW signal of individual MBHBs.

The higher frequency GW from massive binaries could be detected by long-baseline space interferometers such as the Laser Interferometer Space Antenna\footnote{\url{https://www.elisascience.org/}} (LISA) \citep{Danzmann1996, LISA2017, Klein_etal2016}, from the European Space Agency\footnote{\url {https://www.esa.int/}} (ESA), which has recently been adopted and is expected to be launched in the next decade \citep{LISA2024}.
 LISA will be sensitive to GWs with frequencies in the range $100\mu{\rm Hz}$--$100 {\rm mHz}$, which corresponds to MBHBs with total masses between $10^4$ and $10^7~M_\odot$ in the last stages of their evolution.
 LISA is expected to detect black hole mergers up to a {\it redshift} of $z \sim 20$, corresponding to cosmic epochs where the first seeds of SMBHs were formed \citep{AmaroSeoane_etal2023}.

\section{The standard picture of accretion onto massive black hole binaries}
\label{sec:accretion_simple}

MBHBs are located in the center of merged galaxies.
These environments may have sufficient amounts of gas for the MBHB to develop similar features to normal AGNs; in particular, the formation of an accretion flow and (potentially) outflows.

At sufficiently large distances from the center of mass of the system, $r \gg a$, matter from the ambient medium feels the gravitational influence of the binary system as that of a single object with a total mass $M=m_1 + m_2$. The orbiting gas would gradually fall inward by transferring angular momentum outwards while generating heat \citep{Frank_etal2002}.
This process would be primarily driven by viscous dissipation  \citep{Pringle1981} triggered by the magneto-rotational instability (MRI) and the resulting MHD turbulence \citep{BalbusHawley1991_MRI}.

The evolution of the flow may be drastically different depending mainly on the accretion rate and angular momentum supplied at the Bondi radius.
For low accretion rates, i.e., much smaller than the Eddington rate, $\dot{M}_{\rm edd}=1.38 \times 10^{18} (M/M_\odot)~{\rm g~s}^{-1}$, the densities will be low and the gas will form a radiatively inefficient accretion flow (RIAF) \citep{YuanNarayan2014}; here, most of the thermal energy is advected to the central object with large radial velocities and sub-Keplerian orbits.
A similar flow structure arises when the accretion rate is super-Eddington and the photon energy gets trapped in the gas, radiating very little. Given that the sound speed is comparable with the Keplerian orbital velocity in these cases, the accretion flow tends to be geometrically thick. 
We will focus on radiatively efficient flows, which have accretion rates close to the Eddington limit and are more promising for multimessenger astrophysics.
For the reader interested in the RIAF-like regime in binaries, we refer to Ref. \cite{CattoriniGiacomazzo2024} and references therein.

Far from the binary, a luminous accretion flow will resemble a standard thin disk around a point mass in which the disk is cooled by radiating the dissipated energy of accretion into photons \citep{ShakuraSunyaev1973}. As the gas gets close enough to the binary, it will start to feel the effects of the \emph{dynamic and non-axisymmetric gravitational field}, which will significantly modify the flow structure.
The main influence of the binary on the accreting gas is through the positive torque that the rotating potential imprints on the matter \citep{LyndenBell_Kalnajs1972, GoldreichSari2003}. Indeed, the time-dependent quadrupolar potential of the binary does not conserve the energy or the angular momentum of the plasma.
It was first noticed by Artymowicz \& Lubow \cite{ArtymowiczLubow1994} that if this effect dominates the flow's angular momentum evolution, the torque will push the gas away from the binary \emph{carving a low-density cavity in the inner region where torques and viscous stresses balance each other.}
For close to equal-mass binaries, the accretion disk is then truncated at a distance $r_{\rm edge} \sim (2-3) a$ from the center of mass, as shown by many different simulations \citep[e.g.,][]{Farris:2014ApJ, DOrazio13, Noble12}.

The truncated disk orbiting the binary is usually referred to as the \emph{circumbinary disk} (CBD), and it was first suggested to exist around MBHBs by Begelman et al. \cite{Begelman1980}.
The first models on CBDs \citep{Pringle1991} assumed that no accretion would occur within the cavity.
If this were the case, we may conclude that since the luminosity of an accretion disk depends on how close its inner edge is from the gravitating source ($L_{\rm disk} \propto R_{\rm in}^{-1}$; see, e.g., \cite{Frank_etal2002}), accreting MBHBs would be much dimmer than normal AGNs.
This would make it much harder to detect EM counterparts of SMBH mergers, decreasing the multimessenger prospects for these sources.

Fortunately, in more recent works a different and more promising picture has consistently emerged from results from various types of simulations using Newtonian hydrodynamic \citep{MacFadyen2008, DOrazio13, DOrazio16, Farris:2014ApJ, MunozLai16, Moody2019, Duffell2020, Mosta2019, Munoz2019, Munoz2020, Zrake2021, Tiede2020, Derdzinski2021, DittmannRyan2021} with and without viscosity, Newtonian MHD \citep{Shi:2012ApJ}, GRMHD with approximate spacetimes \citep{Noble12, lopezarmengol2021}, and full NR + MHD simulations \citep{Farris:2011PhRvD, Gold_2014a, paschalidis2021minidisk}.
It is now accepted that a significant fraction of the matter in the inner edge of the CBD falls towards the binary system, mainly through thin streams.
Moreover, it was shown that, generally, these streams of matter have enough angular momentum to start orbiting the individual holes forming a pair of circum-single disks, also known as \emph{mini-disks}.

A unique characteristic of CBD accretion of binaries with mass-ratios $q \lesssim 1$ is the presence of a `lump'; this is a \emph{coherent} overdense region ($m=1$ mode) that orbits the binary close to the inner edge of the CBD.
The formation of a lump has been consistently demonstrated through simulations of CBD accretion with different approaches: 2D hydro \citep{MacFadyen2008, Farris:2014ApJ, DOrazio13, DOrazio16, Miranda2017, Moody2019, MunozLai16, DittmannRyan2022}, 3D Newtonian MHD \citep{Shi:2012ApJ, Shi2015} and 3D-GRMHD \citep{Noble12, Gold_2014a, noble2021, lopezarmengol2021}. As we will discuss later, the lump is a very important feature of a CBD because it modulates how the mass flows onto the cavity.

\begin{figure}[ht!]
    \centering
    \includegraphics[width=0.7\textwidth]{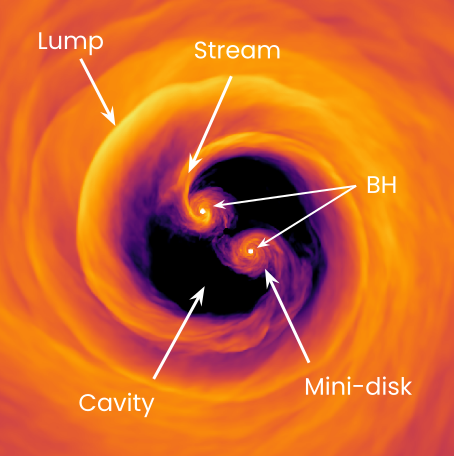}
    \caption{Gas density in the equatorial plane around a quasi-circular equal-mass binary at a separation of $a\sim 20\,M$ from a 3D GRMHD simulations \citep{combi2022minidisk}. The different components of the accretion are indicated through arrows.}
    \label{fig:diagram}
\end{figure}

To summarize, we can draw a schematic picture of `standard' thin-disk accretion onto MBHBs as follows: (a) the matter flows inwards from large distances transporting angular momentum outward through viscous stresses due to MHD turbulence; (b)  binary torques carve an eccentric cavity with a mean radius of $r_{\rm cav} \sim(2-3)a$, which also determines the inner boundary of the truncated CBD;  (c) part of the matter reaching this inner boundary is accreted onto the binary through free-fall thin streams connecting it with the black holes and, potentially, forming mini-disks around each binary component. In Figure \ref{fig:diagram}, we show the various components of the accretion flow onto an MBHB on the orbital plane.

We should remark that this picture is approximately valid for near-equal-mass binaries with an aligned cooled thin disk. Accretion onto either eccentric binaries, misaligned disk, or small mass-ratios $m_{1}/m_{2} \ll 1$, can drastically change the system. If we focus on relatively short binary separations, $a \lesssim 100\,M$, i.e. where the binary is efficiently emitting GWs, quasi-circular, and aligned equal-mass binaries may be a natural outcome of long-term evolution. As the binary shrinks, its initial eccentricity would also be radiated away by GWs \citep{Peters:1964zz}; on the other hand, for unequal mass binaries, the less massive component will capture most of the mass and grow through accretion, driving the system to equal-mass; finally, if the flow is initially misaligned with the orbit, the binary torque may align the gas close to the center of mass. Of course, this all assumes that the gas supply at the Bondi radius remains roughly in a steady state for a sufficiently long time. 

\begin{figure}
    \includegraphics[width=.95\textwidth]{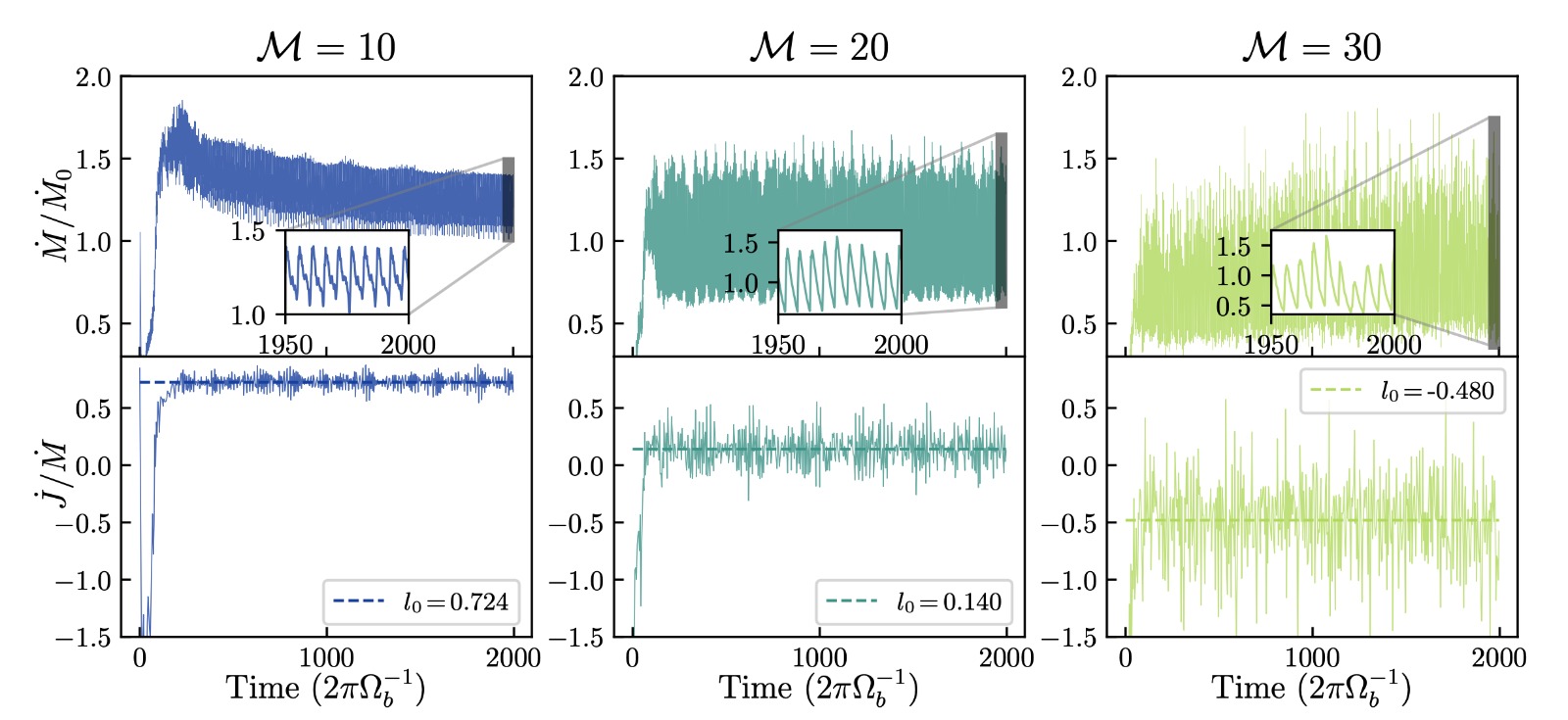}
         \caption{Accretion onto the cavity for different temperatures (Mach numbers) in an equal-mass binary, showing strong modulations, from a 2D VH simulation. Figure from Ref. \cite{DittmannRyan2022}.}
         \label{fig:mdot_ditt}
\end{figure}

Accreting MBHBs present several hydrodynamical quasi-periodicities that might help distinguish them from single AGN \citep{MacFadyen2008, Dorazio2018, Bowen2017}, as shown in the quasi-periodic accretion rate onto the BHs, see for instance Figure \ref{fig:mdot_ditt}.
The light curves of AGNs have large variabilities dominated by red noise, which makes the identification of periodicities extremely challenging \citep{DOrazioCharisi2023}. Accurate models of binary accretion are thus essential to finding a real `smoking gun' in the EM signatures which may lead to detection.

\section{Theoretical models of binary accretion}
\label{sec:modeling_MBHBs}

The accretion flow onto a binary will differ from that with a single black hole only close to the source, where the effect of the binary quadrupole ($\sim r^{-3}$) becomes important relative to the gravitational monopole ($\sim 1/r^2$).
As we go closer to the black holes ($r< 2 a$) and binary separations are small $(a <100M)$, GR becomes important to describe the interaction between the plasma and spacetime.
The gravitational field is non-axisymmetric and non-stationary, providing an additional source of torque and energy, changing how angular momentum is transported in the system.

The physical ingredients for describing binary accretion do not differ greatly from single black hole accretion except that now we must account for the dynamical spacetime, or in Newtonian language, the time-dependent quadrupolar gravitational field. We must then solve Einstein's equations, coupled with the fluid conservation equations and, ideally, radiation transport. On this last point: many simulations to date treat radiation at best with a crude effective cooling function \citep{Noble:2009ApJ, Noble12, Gold_2014a, westernacher2022multiband}, whereas most Newtonian simulations neglect radiation entirely and instead evolve an isothermal gas \citep{duffell2024santa}.

Because the accretion flows coupled with magnetic fields are essentially turbulent and the gravitational potential is time-dependent and non-axisymmetric, analytical models can only treat a few qualitative aspects of binary accretion \citep{Binney1987}.
Even to get qualitative predictions, numerical solutions are needed.  
A big challenge for these simulations is resolving the large dynamic range of time and length scales that govern the system's evolution. Global simulations must simultaneously include the black hole horizon $\sim R_{\rm g}$ and the CBD of size $\gtrsim 3a \gg R_{\rm g}$, as well as resolving time-scales of the order of the orbital period $\sim P_{\rm bin}$ while evolving to much longer viscous timescales for the disk to reach an inflow-outflow equilibrium \citep{LaiMunoz2023}. 
The viscous timescale $t_{\rm visc} \sim r^2/\nu \sim \alpha^{-1} (h/r)^{-2} (r/a)^{3/2} P_{\rm bin}$, can be easily longer than $100 P_{\rm bin}$ at moderate radii.
This implies that simulations must be evolved for hundreds of orbits to reach a steady state independent in some measure from initial conditions. Because of the computational cost, long-term simulations can be unfeasible unless approximations are made.
These will depend on the specific questions one wants to address.
On this aspect, the literature is roughly divided into Newtonian 2D viscous-hydrodynamics (VH) and 3D GR-magnetohydrodynamics, with a few simulations in between studying binaries in 3D Newtonian MHD \citep{Shi:2012ApJ, Shi2015}, 3D Newtonian VH \citep{DittmannRyan2022}, and a few 2D GR-hydrodynamic (GRHD) simulations \citep{mignon2023origin}.
Although 3D GRMHD can capture the correct angular momentum transport and resolve the physics around the black hole horizon, the current computational cost of each simulation is extremely expensive to run for $\gtrsim1000$ orbits.
Two-dimensional VH simulations, instead, are much cheaper (by a factor of $>10$) and can be used to model long-term evolution and explore the parameter space.
However, because VH simulations do not model magnetic fields and cannot resolve the physics close to the black holes, they cannot accurately model the EM emission that comes out from the innermost regions.

In the next subsections, we briefly present the main aspects of each of these two approaches.

\subsection{Newtonian 2D viscous hydrodynamics}

For luminous AGNs, those that we can observe up to very large distances, the accretion disk must be efficiently converting gravitational potential energy in radiation and thus maintaining (relatively) low temperatures \citep{Frank_etal2002}.
If the thermal pressure remains low, the disk must be geometrically thin $H/r \ll 1$.
The vertically integrated 2D hydrodynamical equations on the azimuthal plane $(r, \phi)$ are thus a reasonable zeroth order representation of the system.
In addition, if one is mostly interested in the orbital secular evolution due to accretion, we may assume that the binary separation is large and neglects its evolution through GW emission.
As long as we focus on scales far from the black holes, we can also assume that the gravitational field of the binary is well described by a Newtonian potential of two-point masses of the form
\begin{equation}
    \Phi(t,\vec{r}) = -\frac{m_1}{|\vec{r}-\vec{r}_{1}(t)+\epsilon_{\rm g}|} - \frac{m_2}{|\vec{r}-\vec{r}_{2}(t)+\epsilon_{\rm g}|},
\end{equation}
where $r_{i}(t)$ follows a Keplerian orbit and $\epsilon_{\rm g}$ is the softening gravitational length to avoid numerical singularities in the simulation \citep{DittmannRyan2022}.
Because we are not resolving the black hole horizon explicitly, a subgrid model of accretion is needed to account for the amount of mass and angular momentum that exits the system.
For this purpose, sink terms for mass and momentum are included in the equations to extract the gas that is accreted.
Several methods have been used in the literature, including special techniques to accrete without introducing spurious torques \citep{DittmannRyan2021}.
An alternative solution pursued in early work was instead to excise the inner region entirely, placing an inflow-only boundary condition inside the cavity at a radius $r \sim a$ \citep{MacFadyen2008, Shi:2012ApJ, Noble12, lopezarmengol2021}.
While this alleviates the need for sub-grid source terms, this approach precludes the possibility of gas being recycled from the mini-disks back into the cavity, does not allow simulation of the mini-disks at all, and cannot measure the accretion rates onto the individual black holes.

In the vertically-integrated MHD equations, magnetic stresses due to MRI cannot be self-consistently generated nor sustained due to the necessity of a vertical field \citep{gilbert2003handbook}.
Hence, the source of viscous stresses needs to be approximated with a prescription that relates this to the local properties of the gas.
In the equatorial plane, the relevant stresses operate in the $r\phi$ direction and can be approximated by a kinematical viscosity, $\nu$, which might be fixed directly or indirectly through an $\alpha$-prescription, where $\nu = \alpha c_{\rm s} H \propto \alpha \mathcal{M}^{-2}$.
The 2D Newtonian hydrodynamical vertically integrated equations can be found explicitly in Eqs. (1-3) in \citep{DittmannRyan2021}. The equation of state is usually an isothermal one, $\Pi = c_{\rm s}^2 \Sigma$, where $\Sigma = \int \rho dz$ is the surface density, $\Pi = \int P dz$ is the vertically integrated pressure, and the sound speed $c_{\rm s}$ is chosen proportional to the gravitational potential and Mach number $\mathcal{M}$,  $c_{\rm s}^2 = -\Phi \mathcal{M}^{-2}$.
Other works have used a gamma-law gas equation of state, $\Pi = (\Gamma-1) \Sigma \epsilon$, and the so-called $\beta-$cooling, which adds a cooling function to the equations \citep{Farris:2015MNRAS, Tang2017, WesternacherSchneider_etal2022} and can be used as a proxy for radiated energy. 

Several finite-volume and smooth-particle hydrodynamical codes have been used to evolve the 2D Newtonian equations of the state including the moving-mesh  {\tt  DISCO}, the multi-purpose Cartesian/spherical {\tt  Athena++}, {\tt Pluto}, and {\tt AREPO}, and smooth particle hydrodynamical (SPH) codes like {\tt Phantom}. A code comparison project has recently shown good agreement among these codes \citep{duffell2024santa}. These simulations are sufficiently cheap to run with relatively high resolution for $\sim 1000$ orbits which is needed to reach a steady state on longer viscous timescales. Many crucial insights in binary accretion have been solved using 2D hydro, especially related to binary evolution \citep{LaiMunoz2023}.

Two-dimensional VH Newtonian simulations can capture several sources of angular momentum flux, including most importantly, the gravitational torques. These simulations, however, use a very simplified representation of the internal magnetic stresses present in real systems. Moreover, most of these use fairly unrealistic thermodynamics given by isothermal EOS and neglect all 3D effects. As pointed out in \cite{noble2021}, this means that \emph{VH simulations are never turbulent and show smooth and steady flow}, which are quite different from real AGNs.

\subsection{General relativistic 3D magneto-hydrodynamics}
\label{sec:2}

Because magnetic stresses are the main mechanism of angular momentum transport in accretion disks, 3D MHD simulations are needed to accurately model this aspect of accreting MBHBs.
Close to the black holes, where most of the `viscous' dissipation and emission occurs, a general relativistic description of the gravitational field becomes essential too.
Moreover, if the black holes are spinning, accretion can drive magnetic flux to the ergosphere and launch powerful double jets, which is a purely GR effect.
If we want our emission models from binary black hole accretion to be accurate and realistic, we need to incorporate both MHD and GR. 
This is challenging because the flow around the black holes is still highly coupled with the outer circumbinary accretion disk, which evolves much more slowly. 

The equations of GRMHD are obtained from (a) the conservation of the energy-momentum tensor, (b) the conservation of rest-mass density, and (c) the Maxwell's equations in the ideal MHD limit of infinite conductivity.
These equations must be supplemented with an equation of state, e.g. relating pressure and internal energy.
In addition, the spacetime evolution is coupled to matter through the covariant derivatives in the equations.
The (dynamical) metric of the binary can be obtained by solving Einstein's equations numerically or using some approximation (see below). 

To solve the equations numerically in a computer, these can be cast in a conservative form using an explicit 3+1 decomposition as in the Valencia formulation used by the numerical relativity community \cite{Farris11}, or in a coordinate split as used in the {\tt Harm3D} code \citep{Noble12}.
The coupled equations of motions, following \citep{Noble12}, can be written as
\begin{equation}
\partial_t \bU\left(\prim\right) =                                                                    
-\partial_i \bF^i\left(\prim\right) + \mathbf{S}\left(\prim\right),
\label{eq:cons-form-mhd}
\end{equation}
where $\bP$ are the \textit{primitive} variables, $\bU$ are the \textit{conserved} variables, $\bF^i$ are the \textit{fluxes}, and $\mathbf{S}$ are the \textit{source} terms.
These are given explicitly as
\begin{eqnarray}
\mathbf{P} & := & [\rho, u, \tilde{u}^j, B^{j}] \ ,
\label{primitive-mhd} \\
\bU\left(\prim\right) & := & \sqrt{-g} \left[ \rho u^t ,\, {T^t}_t +
  \rho u^t ,\, {T^t}_j ,\, B^j\right] \ , \label{cons-U-mhd} \\
\bF^i\left(\prim\right) & := & \sqrt{-g} \left[ \rho u^i ,\, {T^i}_t +
  \rho u^i ,\, {T^i}_j ,\, \left(b^i u^j - b^j u^i \right)\right], \label{cons-flux-mhd} \\
\mathbf{S}\left(\prim\right) & := & \sqrt{-g} \left[ 0 ,\,
  {T^\kappa}_\lambda {\Gamma^\lambda}_{t \kappa} - \mathcal{F}_t ,\,
  {T^\kappa}_\lambda {\Gamma^\lambda}_{j \kappa} - \mathcal{F}_j ,\, 0
  \right], \label{cons-source-mhd}
\end{eqnarray}    
where $\rho$ is the rest-mass density, $T^a_{b}$ are the components of the energy-momentum tensor, $g$ is the determinant of the metric, ${\Gamma^\lambda}_{\alpha \beta}$ 
are the Christoffel symbols, $u:=\rho \epsilon$ is the internal energy, $\tilde{u}^j:= u^j -g^{tj}/g^{tt}$ is the velocity relative to the normal spacelike hypersurface, and $B^{j}:= \: ^{*} F^{it}$ is the magnetic field, which is both a conserved and a primitive variable. Finally, $\mathcal{F}_j$ is a contribution to the source term that can account for radiative cooling \citep{Noble:2009ApJ}.

Although relatively less explored than 2D VH simulations, in the last decade several 3D GRMHD simulations have investigated binary black hole accretion \cite{Gold2019}.
These simulations have focused on the late inspiral stage before the merger, starting from separations of $\sim 10-20\:M$ and evolving for a few tens of orbits when the black holes are included in the domain.
3D GRMHD simulations have to deal with two main issues: a) long evolutions to reach a true reliable steady state, which makes them very expensive, and b) accurate modeling of the time-dependent binary spacetime.

A 3D GRMHD simulation of binary black hole accretion must include a dynamic representation of the spacetime.
Depending on the regime of binary evolution, different approaches have been taken in the literature, from PN approximations to full numerical relativity calculations, i.e., numerically solving Einstein's equations coupled with the MHD equations.
We summarize the various approaches below:

\begin{itemize}

    \item \textbf{Weak-field approximation}: For modeling the circumbinary gas flow far away from the black hole, one can use a weak-field PN approximation for the metric \citep{Blanchet:2014av}. This metric is an expansion in powers of $v/c$ and is usually derived in harmonic coordinates. It represents two point particles following PN trajectories and it is valid for binary separations of $a>10\:M$. It has been used in several simulations that focused on the CBD while excising the inner region of the binary \cite{Noble:2009ApJ, noble2021, Zilhao2015}. 

    \item \textbf{Strong-field approximation}: Well before the merger, tidal interactions between the black holes are small and GW effects are mostly relevant for the trajectories of the black holes. A strong field representation that includes black hole horizons explicitly can be built in a variety of ways without numerical relativity (i) using a matching metric, where different approximations are sewed together \citep{Mundim2014}, (ii) using a conformal thin sandwich approximation for binaries on quasi-circular orbits \citep{Farris:2011PhRvD}, and (iii) using a linear superposition of black holes e.g. in Kerr-Schild coordinates \citep{combi2022minidisk}. Simulations using approximations like these are much cheaper than those using full numerical relativity and can be used to evolve the flow near the black holes for separations of $a > 10M$ (however, see also the approximation of \cite{CombiRessler2024} which is valid even at merger).

    \item \textbf{Numerical relativity}: For evolving the binary from small separation to merger, several simulations have performed full GRMHD calculations where Einstein's equations are solved along with the fluid equations \citep{Farris:2012PhRvL, Gold_2014a, Gold_2014b, paschalidis2021minidisk, bright2023minidisc, Ruiz_etal2023}. In this approach, the energy-momentum of matter is usually neglected and the vacuum equations are evolved using a 3+1 decomposition in the Baumgarte--Shapiro--Shibata--Nakamura (BSSN) formalism \citep{baumgarte}. The first GRMHD simulation of a binary black hole with numerical relativity including a CBD was conducted by Farris et al. \cite{Farris:2012PhRvL}, who considered equal-mass black holes and explored disks with and without cooling for $\sim 45$ orbits starting at $10\,M$
\end{itemize}

Most 3D GRMHD simulations have used an ideal gas equation of state, with adiabatic index given by $\Gamma = 4/3$ or $\Gamma=5/3$. There are no simulations, as far as we know, that include proper photon transport in this context. An effective isotropic cooling function was included in several works to cool the disk to a target entropy on timescales related to the local Keplerian time \citep{Farris:2012PhRvL, Gold_2014a, Gold_2014b, Noble12, noble2021, Bowen2019, Bowen17, combi2022minidisk, Gutierrez_etal2022}; various prescriptions for this cooling time were proposed and tested in Ref. \citep{Gold_2014a} using ideas from Ref. \cite{Noble:2009ApJ}. This cooling function is physically ad-hoc but maintains the disk geometry to a fixed aspect-height ratio, similar to the standard isothermal approach of 2D VH but allowing for dynamic cooling and heating. The heat extracted from the system via the cooling function is used then to calculate the radiation output of accretion, see Section \ref{sec:em}.

GRMHD simulations of single black hole accretion usually start with a hydrostatic axisymmetric disk solution which can be constructed around a Kerr black hole using different prescriptions, e.g. with constant specific angular momentum as proposed by Fishbon and Moncrief \citep{Fishbone1976} or with a power-law relationship between the specific angular momentum and angular velocity \citep{DeVilliers2003, chakrabarti1985natural}. This equilibrium is then perturbed by the weak magnetic field threading the disk (plus some pressure perturbations) and a turbulent state is developed. In most GRMHD simulations of binary accretion, the circumbinary disk was initialized following Ref. \citep{chakrabarti1985natural} and setting up the inner edge of the disk at $\sim 3 a$, where the cavity would be approximately located, and the radius of maximum pressure at $\sim 5 a$. In \cite{Farris12} and subsequent work by the same collaboration, a hydrostatic torus solution is initialized on a single black hole spacetime. Differently, in \citep{Noble12} and subsequent work by the same collaboration, the hydrostatic solution was set up in an axisymmetric metric constructed as the azimuthal average of the binary metric. 

To obtain inflow equilibrium on the circumbinary disk, where the system evolves from an artificial equilibrium state to a fully turbulent steady state, 3D MHD simulations must evolve for more than $\sim 100$ orbits, which is challenging when the black holes are included in the simulation grid. For this purpose, Refs. \citep{Bowen2019, combi2022minidisk, avara2023accretion} used a hybrid approach where the circumbinary disk is evolved first, excising a sphere around the binary, until a steady state is achieved at the inner edge, and then starting the simulation with the black holes on the grid. For quasi-circular orbits, this approach is likely justified because the inner part of the binary has an effective ``accretion horizon'' where the matter is trapped and accreted \cite{Tiede2021}.

The magnetic field in all these simulations is set as a single poloidal loop with a maximum strength set by a ratio of fluid pressure and magnetic pressure of $\sim 100$. This leads to a regime of standard accretion, in contrast to a magnetically-arrested disk (MAD)\citep{tchekhovskoy2011efficient}. We notice that the MAD regime has not been investigated for equal-mass binaries that develop a cavity (see however \cite{noble2021} for a discussion on magnetic field effects on the CBD). On the other hand, Ref. \cite{ressler2024black}
investigated the case of a MAD, hot accretion flow around a small mass-ratio inclined binary, finding the precession of the jet due to spin-orbit coupling and periodicities in the mass outflows.
 
In Table \ref{tab:GRMHD_sims}, we collect all the Refs. we are aware of where results from 3D-GRMHD simulations of accreting MBHBs were presented.
We include in the table the numerical approach used to model the dynamical spacetime, the initial data, the grid, and the code used in each paper.
We also discriminate those simulations that included the black holes on the grid (BHOG) from those that excised the inner region, as well as those that included cooling from those that did not include it.

\begin{table}
    \centering
    \begin{tabular}{c|c|c|c|c|c|c}
        Ref. & Spacetime & BHOG & Cooling & Initial Data & Grid & Code \\
        \hline \hline
        
        \cite{Farris:2012PhRvL} & NR & \cmark & \cmark & Hydrostatic torus & Cart. & {\tt IllinoisGRMHD} \\
        
        \cite{Noble12, Zilhao2015, noble2021} & PN & \xmark & \cmark & Hydrostatic torus & Sph. & {\tt Harm3D} \\
        
        \cite{Gold_2014a, Gold_2014b} & NR  & \cmark  & \cmark & Hydrostatic torus & Cart. & {\tt IllinoisGRMHD} \\
        
        \cite{Bowen_etal2018, Bowen2019} & Matching \cite{Mundim2014}  & \cmark  & \cmark & Relaxed CBD \cite{Noble12} & Warped  &  {\tt Harm3D} \\
        
        \cite{lopezarmengol2021} & SKS & 
        \xmark & \cmark & Hydrostatic torus & Sph. & {\tt Harm3D} \\
        
        \cite{combi2022minidisk} & SHPN \cite{Combi2021a}  & \cmark & \cmark & Relaxed CBD  & Warped  & {\tt Harm3D} \\
        
        \cite{avara2023accretion} & Matching & \cmark & \cmark & Relaxed CBD & Multipatch & {\tt PatchworkMHD+Harm3D} \\
        
        \cite{cattorini2021fully, cattorini2022misaligned} & NR & \cmark & \xmark & Uniform gas & Cart. & {\tt IllinoisGRMHD} \\
        
        \cite{Ruiz_etal2023} & NR & \cmark & \xmark & Hydrostatic torus & Cart. & {\tt IllinoisGRMHD} \\
        
        \cite{paschalidis2021minidisk, bright2023minidisc} & NR & \cmark & \xmark & Hydrostatic torus & Cart.  & {\tt IllinoisGRMHD} \\
        
        \cite{Giacomazzo:2012ApJ} & NR & \cmark & \xmark & Uniform gas & Cart. & {\tt Whisky} \\
        
        \cite{Kelly2017} & NR & \cmark & \xmark & Uniform gas & Cart. & {\tt IllinoisGRMHD} \\

        \cite{ressler2024black} & SKS-PN \cite{CombiRessler2024} & \cmark & \xmark & Relaxed MAD disk & Cart. & {\tt Athena++}
                
    \end{tabular}
    \caption{List of references presenting results from GRMHD simulations of accreting MBHBs.
    The abbreviations shown in the table denote the following. NR: Numerical Relativity; SKS: Superposed Kerr--Schild; SKS-PN: Superposed Kerr--Schild + Post-Newtonian; SHPN: Superposed Harmonic + Post-Newtonian; BHOG: Black holes on the gird. \textit{Cart.} refers to simulations using Cartesian mesh refinement with box-in-box grids following the BHs. \textit{Warped} refers to a topological spherical grid with more resolution on the BHs \citep{Zilhao2015}.}
    \label{tab:GRMHD_sims}
\end{table}

\section{Aspects of supermassive binary black hole accretion on the way to merger}

\label{sec:accretion}

There are still many open questions and ongoing work on binary black hole accretion. Some aspects related to circumbinary accretion are shared among other binary accreting systems such as in protoplanetary disks \citep{LaiMunoz2023}; other aspects of the system where GR effects are important are unique to MBHB accretion. Even within the simple 2D VH approximation, the literature has debated fundamental questions such as whether the net angular momentum flux from the gas to the binary is positive or negative.
There is also a huge range of parameter space to explore both for the gas properties and the binary.
Moreover, recent full 3D MHD calculations of CBD accretion that are evolved for $\sim 200$ orbits \citep{noble2021, lopezarmengol2021} show some qualitative differences with respect to 2D VH simulation. 

As we stated before, we are mainly interested in the MBHB problem when the separations are small and the prospects of multimessenger observations are higher. Within this small subset of possible models, we will discuss the following questions:

\begin{itemize}
    \item What are the mechanisms for transporting angular momentum in the system? What drives accretion from the CBD to the black holes?
    \item What are the properties of the circumbinary cavity? 
    \item How does the \emph{lump} form, and what are their properties? What influence do magnetic fields play on this matter?
    \item Does the binary decouple from the CBD?
    \item What are the differences between single black hole disks and mini-disks? What drives accretion in the mini-disks? 
 \end{itemize}

We purposely neglect other important topics, for instance, related to the long-term evolution of the binary and the influence of the gas, misaligned disks, and large eccentricities. These are crucial aspects of binary accretion but in service of focusing the review on results where GRMHD is more relevant, we recommend other reviews such as \cite{LaiMunoz2023} for the interested reader.

\subsection{Angular momentum transport in circumbinary disks}

In steady accretion disks around a point-mass potential, angular momentum transport is driven by internal magnetic stresses generated through the MRI \citep{BalbusHawley1991_MRI, Hawley1991II}.
In the case of binary black hole accretion, this is likely the main mechanism for angular momentum transport far from the binary, but the inner regions will be heavily influenced by gravitational torques and the non-axisymmetric, time-dependent gravitational potential.  In addition to MHD turbulence and gravitational torques, the binary potential can induce spiral waves in the disk, which can also transport angular momentum as they shock and become non-linear \citep{Shi2015}. 
In 2D VH, the internal (viscous) stresses are modeled by an $\alpha$ prescription or a fixed kinematic viscosity. However, other mechanisms of angular transport such as the gravitational torques are captured in 2D VH with some limitations related to isothermal equations of state \citep{miranda2019multiple} and the lack of 3D effects.

Before describing in more detail how these mechanisms operate, we should say a few things about how to measure the different contributions to the angular momentum flux. Because the gravitational potential is time-dependent and non-axisymmetric, angular momentum and energy are not conserved. We can however define a preferred coordinate system on the center of mass where a total angular momentum can be defined even if it is time-dependent (in this frame, the total angular momentum at infinity will be, however, conserved). The procedures to obtain this quantity are different in GRMHD and Newtonian physics. In 2D VH \citep{Miranda2017}, we start defining the Newtonian total angular momentum per unit $r$ as
\begin{equation}
    \frac{dJ_{\rm Newt}}{dr} = \oint r \Sigma (r^2 \Omega) d\phi,
\end{equation}
where $\Omega$ is the angular velocity.

Using the equations of motion, it is possible to derive a balance law as
\begin{equation}
    \frac{d\dot{J}_{\rm Newt}}{dr} =  \frac{d\dot{J}_{\rm adv}}{dr} - \frac{d\dot{J}_{\rm visc}}{dr} + \frac{dT_{\rm Newt}}{dr}.
\end{equation}
which can be integrated in $r$ to yield the total angular momentum flux
\begin{equation}
    \dot{J}_{\rm Newt}=  {\dot{J}_{\rm adv}}- {\dot{J}_{\rm visc}} + {T_{\rm Newt}}
\end{equation}
where ${\dot{J}_{\rm adv}}$ is the advected angular momentum flux, $- {\dot{J}_{\rm visc}}$ the angular momentum flux due to internal viscosity and ${T_{\rm Newt}}$ is the binary torque, which is expressed more naturally as a torque density:
\begin{equation}
    \frac{dT_{\rm Newt}}{dr} = \int r \Sigma \partial_{\phi} \Phi d\phi
\end{equation}

In GR, energy and angular momentum cannot be locally localized and gauge ambiguities may arise when we try to find balance laws for fluxes. One could use for instance pseudo-tensors such as Landau-Lifshitz's or quasi-local surface tensors such as Brown-Yorks's, both of which depend on specific gauge (or frame) choices \citep{szabados2004quasi}. Despite these gauge dependencies, being able to formulate balance laws can give us crucial information about how the system transfers angular momentum. 

To derive a balance law for angular momentum transport we can start from a current that we can define as $j^a = T^{ab} \phi_{b}$, where $\phi$ is the azimuthal direction. If $\phi_{a}$ were a Killing vector, then angular momentum would be conserved. In the general case, we end up instead with $\nabla_{a} j^{a} = T^{ab} \nabla_{a} \phi_{b}$. As shown in Refs. \cite{Farris:2010PhRvD}
 and \cite{Noble12}, this quantity can be integrated to obtain the following balance law:
\begin{equation}
    \frac{d^2J}{dtdr}= \frac{dT}{dr} - \lbrace \mathcal{F}_\phi  \rbrace - \partial_r \lbrace M^r_\phi \rbrace - \partial_r \lbrace R^r_\phi \rbrace - \partial_r \lbrace A^r_\phi \rbrace,
\end{equation}
where braces indicate surface integrals,  $M^r_\phi=2p_{\rm m} u^r u_\phi - b^r b_\phi$ denotes the Maxwell stresses, and  $R^r_\phi$ and $A^r_\phi$ denote the Reynolds and advection stresses, respectively, which sum to the hydrodynamic part of the stress-energy tensor: $(T_{\rm H})^r_\phi \equiv R^r_\phi + A^r_\phi = \rho h u^r u_\phi$. The spacetime torques are described in a relativistic framework as
\begin{equation}
    \frac{dT}{dr} = \int T^{ab} \partial_{\phi} g_{ab} \sqrt{-g} d\theta d\phi.
\end{equation}

 If $\phi$ is chosen as the orbital azimuthal direction, we recover the Newtonian limit at large distances \citep{Noble12}. We can integrate over $r$ these quantities to obtain the total angular momentum flux as before, although most of the existing literature in GRMHD works directly with $d^2J/drdt$ which is in a sense a more local quantity (a torque density) \citep{Noble12, lopezarmengol2021, noble2021}

These quantities are usually averaged in time to measure the net transport.  Once the disk reaches an inflow equilibrium, the total angular momentum flux does not depend on $r$. This is important to understand whether the gas delivers or extracts angular momentum from the binary. Several 2D VH simulations (and 3D VH simulation \citep{Moody2019}) have recently found that for equal-mass binaries, $\alpha=0.1$, and $h/r=0.1$, the angular momentum delivered to the binary is positive $\ell_0 = \langle \dot{J}_{\rm bin} \rangle/  \langle \dot{M}_{\rm bin} \rangle = 0.68 a \Omega_{\rm bin}$, where $\langle \dot{J}_{\rm bin} \rangle \equiv  \langle \dot{J}_{\rm Newt} \rangle$ is the angular momentum delivered to the binary.
A sufficiently large $\ell_0$ (e.g., greater than $3/8$ for circular equal-mass binaries) determines that the orbital separation will increase as the system accretes.
These results contradict the early 3D MHD Newtonian calculations of \cite{Shi:2012ApJ}, which might be attributed to short evolution times $100 P_{\rm bin}$ in this simulations; it is an open question what happens with longer 3D MHD evolutions. On the other hand, a disk with higher Mach numbers (lower $h/r$) seems to extract angular momentum from the system \citep{DittmannRyan2022}. The torque delivered to the binary is also tied to the eccentricity evolution of the orbit; recent 2D VH simulations found that equal-mass binaries tend to evolve towards an eccentricity of $\sim 0.4$ \citep{Siwek_etal2023, d2021orbital, Zrake2021}.

Hydrodynamic simulations cannot capture the true mechanism for the viscous stresses: MHD turbulence. Therefore, they need to include it with an ad-hoc prescription such as the $\alpha$-prescription, $\nu = \alpha c_{\rm s}^2 \Omega_{\rm K}^{-1}$ \citep{ShakuraSunyaev1973}), or by considering a constant viscosity coefficient $\nu$. Some studies have explored the impact of viscosity on the properties of the accretion disk \citep{DittmannRyan2022}. Spiral density waves generated by the binary, usually exerted as a single-arm $m=1$ density mode propagates through the disk but they seem to carry little angular momentum (\cite{Shi2015}; notice, however, the dynamics of these spiral waves are very sensitive to thermodynamics \cite{miranda2020planet}).

\begin{figure}
 \begin{minipage}[c]{0.35\textwidth}
         \caption{Time-averaged torque density as a function of radius for a CBD simulation, caused by Maxwell stresses (red), Reynolds stresses (green), and advected angular momentum (gold).
         The gravitational torque density is shown in blue, while the radiation losses are shown in cyan. The black line shows the total angular momentum flux. Figure from Ref. \cite{Noble12} with permission.}
         \label{fig:stresses}
    \end{minipage}\hfill
    \begin{minipage}[c]{0.6\textwidth}
    \includegraphics[width=0.95\textwidth]{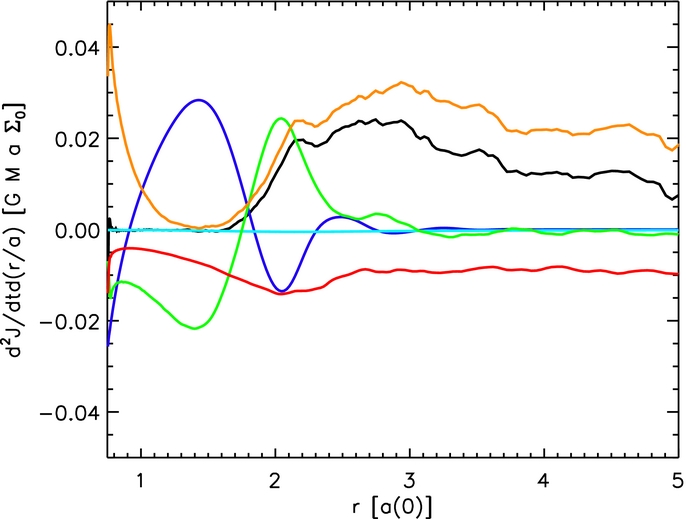}
 \end{minipage}\hfill
\end{figure}

There are very few long-term 3D MHD simulations of CBD accretion \citep{Noble12, Shi:2012ApJ, Shi2015, noble2021, lopezarmengol2021}. Shi et al. \cite{Shi:2012ApJ} conducted the first 3D-MHD simulation of circumbinary accretion onto MBHBs using the {\tt ZEUS} code \citep{StoneNorman1992a, StoneNorman1992b, HawleyStone1995}.
Their study focused on equal-mass black holes at a fixed separation $a$.
They found that the Reynolds stresses are larger than Maxwell stresses near the binary. This finding is consistent with previous research on protoplanetary disk simulations (e.g., \cite{Winters_etal2003}).
While Shi et al. \cite{Shi:2012ApJ} focused on large binary separations, where orbital evolution is negligible, Noble et al. \cite{Noble12} explored the inspiralling regime by including a PN metric and taking into account the shrinkage of the orbit.
These simulations were conducted using the code {\tt Harm3D} with a PN approximation for the spacetime metric.
They traced the evolution of an equal-mass binary from an initial separation of $20M$ to $\sim 8M$, where the PN approximation becomes too inaccurate.
In Figure \ref{fig:stresses}, we show the time-averaged contributions to the angular momentum change in the simulation named \texttt{RunSE} in Ref. \cite{Noble12}.
The figure reveals that binary torques dominate in the gap region $a \lesssim r \lesssim 2a$, and the angular momentum is delivered to the fluid, which shows large and positive Reynolds stresses at $r\sim 2a$.
On the other hand, Maxwell stresses dominate the internal stresses in outside regions, as expected.
A similar analysis has been done in later works \citep{noble2021, lopezarmengol2021}.

Additionally, Noble et al. \cite{noble2021} investigated the internal stress dependence on the mass ratio $q$ and demonstrated a weak correlation. The primary observable effect is the inward shift of the gravitational torque for lower mass ratios. Using 2D VH, Dittman \& Ryan \citep{DittmannRyan2021} showed that the gravitational torque becomes very weak at low mass ratios with $q\sim 0.02$, promoting the outspiral of the binary orbit.

\subsection{Inner parts of a circumbinary disk: the cavity and the lump}

The balance of gravitational torques and viscous stresses gives rise to an inner truncation of the CBD and the formation of a low-density cavity surrounding the black holes. For a disk weakly perturbed by a binary potential, this inner edge is determined by the outer Lindblad resonance \citep{Goldreich1979} which is located at $r \sim 1.3 a$. All recent numerical simulations of binary accretion have shown that the cavity is usually at a larger radius of $r_{\rm trunc} \sim (2-3)a$ \citep{Milosavljevic2005}. The specific shape of the cavity depends on viscous stresses and thermodynamics of the disk. For instance, Ref. \cite{DittmannRyan2022} showed that low viscosities and high Mach numbers lead to an increase in the size of the cavity. In particular, high Mach numbers show large surface density in the inner parts of the disk \citep{Tiede2020}.

Simulations show that the binary torque can induce eccentricity in the cavity. We can define the disk's inner eccentricity (sometimes called lopsidedness) as:
\begin{equation}
    \epsilon \equiv \left \langle \frac{ \left| \langle \rho u^r e^{i\phi} \rangle_\phi \right| }{ \langle \rho r u^\phi \rangle_\phi} \right \rangle_{(r,t)},
\end{equation}
where $\rho$ is the rest-mass density, $u^\mu$ are the components of the fluid four-velocity, and the brackets indicate averages in the subscripted variable. 2D hydrodynamical studies have found that the CBD's eccentricity increases with the mass ratio of the binary from $q \gtrsim 0.05$ up to $q\sim 0.4$, above which it remains roughly constant at $\epsilon \approx 0.1$ \citep{Farris:2014ApJ}.
Simulations in Ref. \citep{lopezarmengol2021} have found that the cavity is less eccentric, stabilizing at  $\epsilon \gtrsim 0.05$ for equal-mass binaries, though Ref. \cite{noble2021} found values about $50 \%$ smaller. Given the large turbulence in MHD disks, the lower value can result from the damping of coherent motions and resonances. However, it can also be that these simulations did not reach a full steady state in terms of this variable \citep{noble2021}. 

So far, we discussed average properties of angular momentum transport but we have not discussed in what way matter actually leaves the inner edge of the CBD and is accreted to the binary. There must be some time-dependent process that allows matter in the inner edge to lose enough angular momentum to be accreted. We know that once a binary component gets closer to the CBD, a thin ballistic stream is pulled from the CBD to the black hole. As shown in Ref. \cite{Shi2015} using 3D MHD with an excised binary, this stream still has too much specific angular momentum to be accreted and is instead flung back to the CBD; the authors argue that the stream will shock against the CBD wall and lose enough angular momentum to be accreted.
In a recent paper, Tiede et al. \cite{Tiede_etal2022} used tracer particles in a 2D VH simulation and argued, on the contrary, that this process is mainly gravitational, driven by tidal stressing (although shocks can still play a role).

For equal-mass binaries on low eccentricity orbits and with cool accretion flows, the inner edge of the cavity develops a strong coherent density structure known as the \emph{lump} \citep{Miranda2017, Noble12, Shi:2012ApJ}. This structure orbits the inner edge of the disk at a Keplerian orbit, given by $P_{\rm lump} \sim 4-5P_{\rm bin}$ for cases with $h/r \sim 0.1$ . The lump is a key feature of the system because it modulates the accretion onto the binary \citep{MacFadyen2008, Noble12, Shi:2012ApJ, DOrazio13, DOrazio16, Farris:2015MNRAS, Zilhao2015, MunozLai16, Miranda2017, Moody2019, Munoz2020, Duffell2020, DittmannRyan2021, lopezarmengol2021}. The modulated accretion depends on the cavity eccentricity as well, which appears to be different in 3D MHD and 2D VH \citep{lopezarmengol2021}. However, most simulations show that the lump periodicity is the strongest periodicity in the CBD accretion. Accretion in this case happens when one of the binary components gets very close to the lump and a large portion of the fluid mass is pulled to the binary.

The formation of a lump requires a mechanism to pile up matter on the inner edge of the CBD in a coherent way.
As the gas falls from the CBD's inner edge toward the cavity, it is thrown back to the disk by the positive gravitational torque exerted by the binary \citep{Shi:2012ApJ, DOrazio13}; this stream of gas hits the disk's inner edge and creates a shockwave that compresses the fluid and produces an overdensity moving with the disk.
This occurs coherently since the orbital frequency at the CBD's inner edge ($r_{\rm edge} \sim 2.5 a$) is a multiple of the orbital frequency ($\Omega({r_{\rm edge}}) \sim \Omegabin/4)$.
Ref. \cite{Munoz2020b} showed that due to the strong density gradient a long-lived $m=1$ mode can exist freely on the inner edge of the CBD\footnote{Because MHD turbulence usually puts more power on longer wavelengths, small azimuthal densities are naturally preferred. However, this strong breaking of symmetry should be further investigated.}.
In its asymptotic state, the lump spans an angular extension of $\pi/3 \lesssim \delta \phi_{\rm lump} \lesssim \pi$ and a radial extension $\Delta r_{\rm lump} \lesssim a$.

Other processes that may contribute to sustaining the lump are the so-called \emph{stream reabsorption} and \emph{lump feeding}.
The first occurs because the binary peels off more matter when the lump is close to the pericenter; then, this stream is thrown back with strong torque and hits the CBD still within the lump since $\delta \phi_{\rm lump}$ can be quite large.
The latter process occurs when the lump approaches the apocenter and its velocity reaches a minimum, while reverse streams are hitting the disk at other phases of the orbit with larger orbital velocities, forming mini-lumps.
These new smaller lumps can catch up with the main lump and join it \citep{Shi:2012ApJ}.

The lump seems to form more easily in 2D hydrodynamical simulations because $\alpha$-viscosity naturally implies little internal stresses in a coherently orbiting structure such as the lump, making it less likely to be pulled apart.
In support of this concept, Ragusa \cite{Ragusa2017} found that the vorticity component perpendicular to the disk plane shows a significant local minimum inside the lump.
This results in the $\alpha$-viscosity effectively emulating the role of MHD physics in promoting the growth and persistence of the lump in CBDs. We should note, however, that the lump seems to even appear in 2D inviscid simulations; \cite{mignon2023origin} attribute the formation of the lump in this case to a Rossby-wave type instability related to vorticity. In other words, there might be different paths to creating the lump in a binary.

Finally, colder and less viscous disks give rise to larger cavities. This produces a longer period for the lump orbit and thus for the accretion rate modulation into the cavity.
Ref. \cite{Ragusa2016} found $t_{\rm lump} \sim 5P_{\rm bin}$ ($t_{\rm lump} \sim 4P_{\rm bin}$) for $h/r = 0.04$ ($h/r = 0.1$), and Ref. \cite{Ragusa2020} reported modulations at $(7-8) P_{\rm bin}$ from a cavity of radius $r \approx 3.5a$, while Ref. \cite{DittmannRyan2022} found that the cavity radius shrink from $3a$ to $2a$ as the viscosity $\nu$ increased from $0.0005$ to $0.008$.

\vspace{0.3cm}
\noindent \emph{What is the influence of magnetic fields on the lump?}

When binary torques throw matter back to the CBD, this also carries with it magnetic field flux.
This flux may increase the magnetic field strength in the region where it shocks the CBD, namely the lump.
To form a lump, the implied magnetic field strength growth rate in the lump must be lower than the density growth rate.
Noble et al. \cite{noble2021} conducted a thorough analysis of this process and found that the magnetic field carried by the torqued streams tends to have opposite polarity to that in the CBD where the stream hits.
This results in large-scale magnetic reconnection, which dissipates the field into heat.
The timescale for this dissipation can be estimated as the period between successive lump-black hole interactions, which is known as the beat period:
\begin{equation}
    t_{\rm diss} \sim \frac{2\pi}{\Omega_{\rm beat}} \approx \frac{2\pi}{2(\Omega_{\rm bin}-\Omega_{\rm lump})} \sim 0.7 P_{\rm bin}.
\end{equation}
On the other hand, the magnetic field is replenished in the lump from the far CBD with the advection timescale, defined as
\begin{align}
    t_{\rm adv} \sim \frac{\Delta r_{\rm lump}}{v_r} \approx \frac{\Delta r_{\rm lump} r_{\rm edge} \Omega_{\rm K}(r_{\rm edge})}{W^r_\phi} \\ \sim
    5 P_{\rm bin} \left( \frac{\Delta r_{\rm lump}}{0.1a} \right) \left( \frac{r_{\rm edge}}{2.5a} \right)^{1/2} \left( \frac{a}{20M} \right)^{-1} \left( \frac{W^r_\phi}{10^{-4}} \right)^{-1},
\end{align}
where we have used the fact that the radial velocity is $v_r \approx W^r_\phi / [r \Omega_{\rm K}(r)]$, with $W^r_\phi$ the ($r\phi$)- component of the Maxwell stress tensor, which is the dominant contribution to the accretion in this region (see Sec. \ref{sec:accretion}).
A lump develops provided $t_{\rm diss} < t_{\rm adv}$, which is indeed true case for typical parameters of binaries in the inspiral regime.
MRI can also operate to increase the magnetic field but this occurs on an even longer timescale: $t_{\rm MRI} \sim P_{\rm lump} \gg t_{\rm diss}$.

The formation of a lump in a binary system is significantly influenced by the mass ratio between the two black holes.
A lower mass ratio results in a lower quadrupole moment, which reduces the strength of the gravitational torques.
This, in turn, reduces the efficiency of the throwing back of matter that feeds the lump.
Additionally, the magnetic field dissipation at the lump will decrease because less magnetic flux will return to the disk.
\begin{figure}
    \begin{minipage}[c]{0.7\textwidth}
    \includegraphics[width=0.95\textwidth]{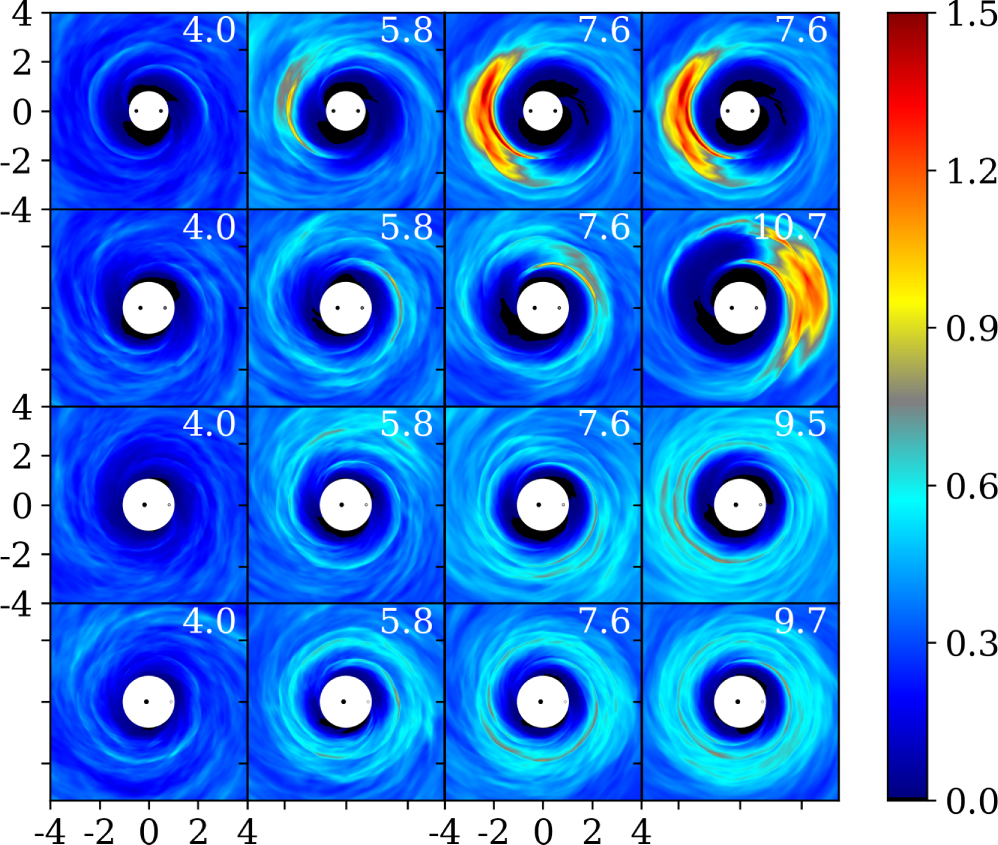}
 \end{minipage}\hfill
 \begin{minipage}[c]{0.29\textwidth}
         \caption{Colormaps of surface density at four different times indicated by the number in the upper-right corner of each image (in units of $10^4M$). The surface density is shown in units of the maximum surface density in the initial state of the disk for the CBD run. From top to bottom, runs with decreasing mass ratio are shown: $q=1,1/2,1/5,1/10$. Figure from Ref. \cite{noble2021} with permission.}
         \label{fig:cbd_lump}
    \end{minipage}\hfill
\end{figure}

In their study, Ref. \cite{noble2021} surveyed the impact of mass ratio (1:1, 1:2, 1:5, and 1:10), and of varying amounts of mass and magnetic flux on the accretion disk. 
Their findings revealed that a significant lump formation requires a sufficiently large mass ratio, as can be seen in Figure \ref{fig:cbd_lump}.
Specifically, the relative amplitude of the overdensity decreases as the mass ratio diminishes, ultimately vanishing somewhere between 1:2 and 1:5.
Regarding the magnetic flux series, they concluded that, whereas the amount of mass in the disk has little influence on the lump properties, the amount of magnetic flux plays a significant role in perturbing the CBD: even a modest, ordered magnetic field was enough to significantly delay the lump formation.
In addition, a greater magnetization weakens any modulation of the light output.

\subsection{Decoupling of the circumbinary disk from the binary}

Once the binary system reaches a sufficiently short distance ($a \lesssim 10^{-3}$ pc), its evolution starts to be dominated by the loss of energy through the emission of GWs.
The timescale for such evolution can be estimated using the post-Newtonian (PN) approximation \citep{Peters:1964zz}
\begin{align}
    t_{\rm GW} := \left( \frac{\dot{a}}{a} \right)^{-1} = \frac{5}{256} \frac{c^5}{G^3} \frac{a^4}{M^2 \mu} \sim 140 \frac{(1+q)^2}{4q} \left(\frac{a}{30 R_{\rm g}} \right)^4 M_7~{\rm days},
\end{align}
where $R_{\rm g}:=GM/c^2$ is the gravitational radius associated with the total mass of the system $M$, $\mu:=qM/(1+q)^2$ is the reduced mass of the system, and $M_7= M/(10^7 ~M_\odot)$.
On the other hand, the accretion timescale for matter in the inner edge of the CBD, assuming it evolves due to viscous stresses, is given by
\begin{equation}
    t_{\rm vis} :=  \left( \left| \frac{v_r}{r} \right|_{r_{\rm in}} \right)^{-1} \sim \frac{2 r_{\rm in}^2 }{ 3 \nu} \sim 30 \left( \frac{r_{\rm in}}{2.5a} \right)^{3/2} \left( \frac{a}{30 R_{\rm g}} \right)^{3/2} \left( \frac{h/r}{0.3} \right)^{-2} \left( \frac{\alpha}{0.15} \right)^{-1} M_7 ~ {\rm days}
\end{equation}
where we have used the $\alpha$ prescription for the kinematic viscosity, $\nu = \alpha H c_{\rm s} \approx \alpha (h/r)^2 l_{\rm K}$, and $l_{\rm K}:= r^2 \Omega_{\rm K}$ is the Keplerian angular momentum.

\begin{figure}
    \begin{minipage}[c]{0.44\textwidth}
         \caption{Timescale and rates for binary orbit evolution due to GW emission and viscosity-driven accretion in the CBD. {\it Top panel:} Timescale-based criterion. {\it Bottom panel:} Velocity-based criterion. Black markers indicate the decoupling separation for each mass-ratio.}
         \label{fig:timescales}
    \end{minipage}\hfill
    \begin{minipage}[c]{0.55\textwidth}
    \includegraphics[width=0.95\textwidth]{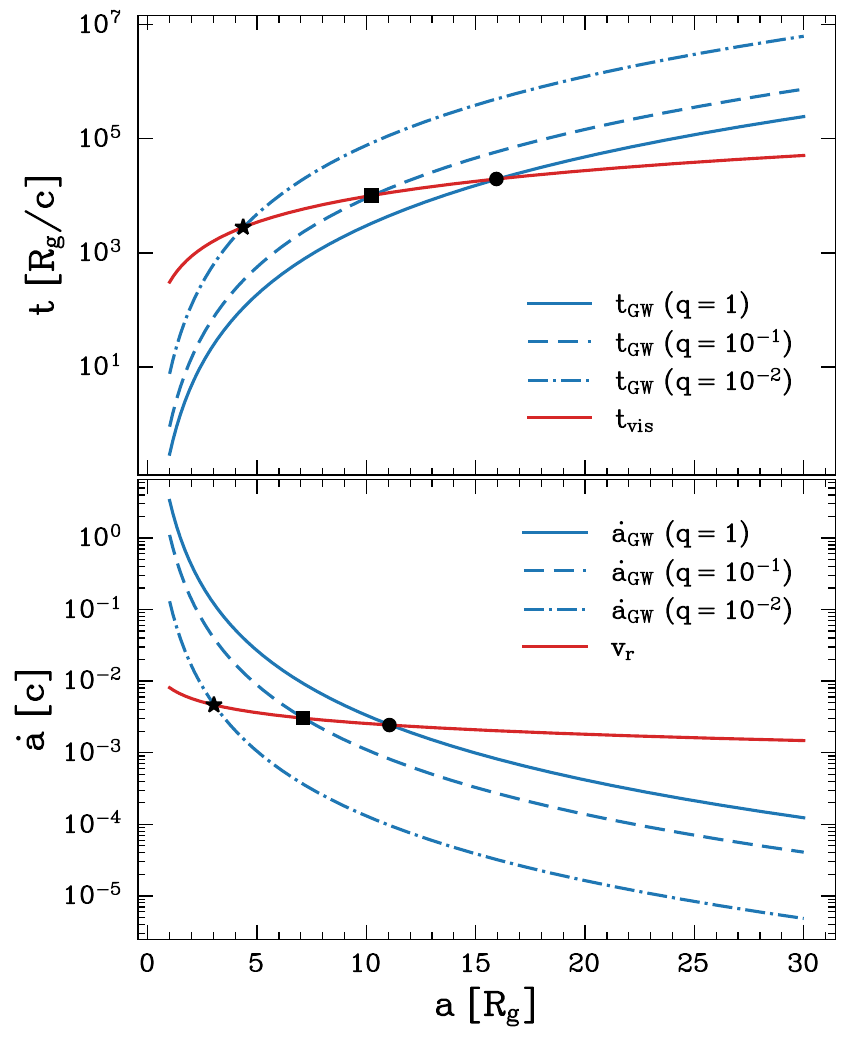}
 \end{minipage}\hfill
\end{figure}

One could predict that when the binary reaches a separation such that $t_{\rm GW} \lesssim t_{\rm vis}$, the shrinking of the orbit becomes fast enough so that the CBD will \emph{decouple} from the binary evolution, and the accretion rate would largely decrease after that.
These two timescales become equal at a separation
\begin{equation}
    a_{\rm d}^{(1)} \sim 16 R_{\rm g} \left( \frac{r_{\rm in}}{2.5a} \right)^{3/5} \left( \frac{h/r}{0.3} \right)^{-4/5} \left( \frac{\alpha}{0.15} \right)^{-2/5} \left( \frac{4q}{(1+q)^2} \right)^{2/5}.
    \label{eq:a_dec}
\end{equation}
Nevertheless, this criterion was found to overestimate the decoupling separation and simulations show a better agreement  \citep{DittmannRyan2023, Milosavljevic2005, Hayasaki07, MacFadyen2008, LiuSLS10, Tanaka10, Kocsis:2011dr, Yunes:2011ws} with a velocity-based decoupling criterion \citep{ArmitageNatarajan2002}, namely, one obtained by equating $\dot{a}_{\rm GW}$ with $v_r|_{r_{\rm edge}}$.
This gives $a_{\rm d}^{(2)} \sim (r_{\rm in}/a)^{-2/5} a_{\rm d}^{(1)} \approx 0.7 a_{\rm d}^{(1)}$, where the last equality holds for $r_{\rm in} \approx 2.5$.
Figure \ref{fig:timescales} shows the two timescales for three different mass-ratios and using the fiducial values in Eq. \ref{eq:a_dec} above.

The expressions above show that the decoupling occurs at closer separations for high CBD's viscosity coefficient and height scale of the disk and for lower mass ratios of the binary.
The dependence of the decoupling separation with $\alpha$ motivates the self-consistent treatment of accretion through MHD simulations, where viscous stresses naturally arise due to turbulence generated by the MRI \citep{BalbusHawley1991_MRI}.
Noble et al. \cite{Noble12} performed the first GRMHD simulations in the inspiralling regime, reaching separations when decoupling is predicted to occur: $\sim (10-20)R_{\rm g}$.
They found an average $\alpha$ parameter of $\approx 0.2$.

This simple picture described above is more complicated in a realistic scenario due to several factors, such as the non-steady behavior or the presence of the lump, but overall simulations show an agreement with the analytical estimations.
Ref. \cite{Noble12} found that after the theoretically predicted decoupling, a significant amount of mass still follows the binary’s inspiral within the gap region, and the accretion rate is still $\sim 70 \%$ the one in the steady state.
Ref. \cite{avara2023accretion} also found that the binary sustains large accretion rates throughout the inspiral and this is only modestly reduced after decoupling.

The decoupling between the CBD and the binary triggers the following important question:
\emph{can MBHBs shine during the merger?} Since most of the radiation from accretion flows comes out very close to the gravitating source, a CBD decoupling from the binary early on may deplete the inner cavity of gas leading to an EM-dim merger.
With the same argument, the falling back of the CBD matter after the merger can give rise to a rebrightening and a delayed post-merger EM counterpart, with a delay
\begin{equation}
    t_{\rm rb} \sim t_{\rm vis}(a_{\rm d}) \sim 6 \left( \frac{r_{\rm in}}{2.5a} \right)^{3/2} \left( \frac{h/r}{0.3} \right)^{-2} \left( \frac{\alpha}{0.15} \right)^{-1} M_7 ~ {\rm days}.
\end{equation}

On the other hand, if the mini-disks are present and they remain healthy after decoupling, there might still be a dense enough environment during the merger to produce EM emission.
GRMHD simulations of mini-disk accretion show an upper bound of mass in the cavity of $M_{\rm cav} \lesssim 2 \times 10^{28} M_7^2 \left( \dot{M}/\dot{M}_{\rm Edd} \right)~{\rm g}$ \citep{Gutierrez_etal2022}.
Nevertheless, this matter would be accreted before the merger if, with an equivalent argument as before, the accretion timescale for the outer boundary of the minidisks is lower than the GW shrinking timescale after decoupling.
In such a scenario, the merger itself could be rather empty of gas and proceed without an EM counterpart, for instance, showing a decrease in X-rays \citep{Krauth_etal2023}.

\subsection{Mini-disks}
\label{sec:minidisks}

Mini-disks (sometimes referred to as circumsingle disks) are rotationally supported accretion flow structures formed around each individual black hole. In a binary system with an external gas supply, the properties of mini-disks are determined strongly by tidal effects and thus by the mass ratio and, once we include GR, orbital separation, and spin. 

A key quantity to characterize a mini-disk is the Hill sphere (or truncation) radius which determines at what point the gravity of the companion disrupts stable orbits. As with the CBD, we can distinguish two regimes of small $q=M_{1}/M_{2} \ll 1$ and near unity $q\sim 1$ mass-ratios. When $M_2 \ll M_1$, a simple Newtonian analysis shows that the truncation radius, defined as the radius where centrifugal force and gravitational force of the companion balance each other, is given by
\begin{equation}
    r_{\rm H} = (q/3)^{-1/3} a
\end{equation}

For small mass-ratio systems and low inclination angles with respect to the disk, the secondary would be embedded in the disk of the primary, e.g. a stellar-mass black hole in an AGN disk. The bounded flow will then be determined by the minimum between the Bondi radius, and the Hill sphere.

For $q\sim 1$, Paczynski \cite{Paczynski:1977} showed with a simple numerical experiment using a pressureless gas that the expression
\begin{equation}
    r_{\rm H} \sim 0.27 q^{-0.3} a,
\end{equation}
can fit the truncation radius fairly well. This has been verified in both 3D MHD and 2D VH simulations \citep{Bowen2017, Bowen2019}. Notice that for large separations, the Newtonian approximation allows us to scale everything with the separation $a$; this scaling is broken when the size of the black holes becomes important. At small separations, the radius of the innermost stable circular orbit (ISCO) is important to determine the size of the mini-disk. The mini-disk will extend from $r_{\rm H}$ to $r_{\rm ISCO} \sim 6\:M$ and will exist as a proper disk as long as $r_{\rm H}/r_{\rm ISCO} >1$, which for non-spinning black holes requires $a > 20 M$ (see below).

\subsubsection{Mass supply and angular momentum flux in mini-disks}

Due to tidal forces and periodic mass supply, mini-disks can behave quite differently than single, steady-state, black hole disks. How mini-disks are formed, in what regime, and how similar they are to single black hole disks are crucial questions to understand what sort of EM signatures we obtain from these systems. Before hydrodynamical simulations, it was unclear whether steady mini-disks could form at all \citep{ArtymowiczLubow1994}. For instance, 
 \cite{tanaka2013recurring} argued that the gas from the elongated stream falling to the black hole will be shock-heated, enhancing stresses in the incipient mini-disk. The large viscosity induced by this would then deplete the mini-disk in short timescales and produce quasi-periodic flaring emission similar to a tidal disruption event. However, as pointed out by Refs. \cite{Roedig2014, Miller:2013ApJ}, the main effect of angular momentum transport in a small disk structure is to viscously spread the gas until the mini-disk reaches the Hill sphere; only then does the gas proceed to be viscously accreted to the black hole. This picture seems confirmed by simulations. Quasi-periodic emission could arise, however, at small binary separations, as we discuss below.

In a mini-disk, contrary to single black hole disks, there may be different sources of angular momentum flux: spiral shocks due to circumbinary stream, tidal-induced shocks by the companion, MRI-driven stresses, and large-scale motions of matter advected onto the black holes \citep{avara2023accretion, RyanMacFadyen17}. The relevance of each contribution is a function of radius (from the black hole) and determines the inflow time of the mini-disk. The inflow time can be defined as $t_{\rm in} \sim \int dr/v^r$, where $v^r$ is the radial velocity of the gas in the frame of the black hole and the integral is computed from $r_{\rm H}$ to $r_{\rm ISCO}$. The binary separation, and more weakly the spin of the black hole, determine the size of the mini-disk and the inflow time. Mini-disks in large-separation binaries are likely to be healthy, similar to single black hole accretion disks near the black hole, while in small separation binaries, the inflow time becomes shorter than the accretion period, driving the system out of a steady state. We discuss this distinction in the next section.

\subsubsection{Mini-disks in and out of equilibrium}

For large separations, the mini-disk size is big and the inflow time is longer than the accretion periodicity from the CBD. In that case, the mini-disks can reach a steady state, where internal magnetic stresses dominate angular momentum transport. We should notice, however, that there are no self-consistent calculations in 3D GRMHD to quantify this in the regime where separations are $a > 20\,M$.
Two-dimensional Newtonian VH simulations are scale-free and can model this large separation regime (as long as the sink that determines the accretion onto the point masses acts on a small region). These simulations have shown that the inflow time is sufficiently long to form coherent structures: the mini-disks never fully deplete and variations in the mass are small \citep{westernacher2024eccentric, Farris:2014ApJ}. In an interesting numerical experiment, Westernacher-Schneider et al. \cite{westernacher2024eccentric} showed that if the CBD is removed abruptly from an equal-mass binary, the mini-disks take $\sim 20$ binary orbits to deplete half their mass. They also show that the mini-disks can grow in eccentricity due to resonant instabilities. On the other hand, in analogy to cataclysmic variables (e.g. Refs. \citep{Ju16}) Ref. \cite{RyanMacFadyen17} analyzed spiral shocks in 2D GRHD simulation of a single mini-disk in the corotating frame, showing that even in the absence of viscosity, tidal shocks induce accretion onto the black holes.

For small binary separations, tidal effects are important, mini-disks are smaller, and the inflow time becomes comparable to accretion timescales from the CBD.
In that case, the Hill sphere of the mini-disk approaches the ISCO, and the spin of the black holes becomes important to determine the size of the mini-disk.
To get an estimate of when the mini-disks start being out of equilibrium, we can calculate the instantaneous inflow time of the mini-disk as $t_{\rm in} \sim r / v^r \sim  (2/3) (1-\sqrt{r_{\rm ISCO}/r})\,r^2/\nu$, assuming a vertically-integrated 2D stationary disk where the stresses go to zero at the ISCO. Taking $r=r_{\rm H}$ and considering $\nu = \alpha c_{\rm s}^2 \Omega^{-1} \sim \alpha (h/r_{\rm H})^2 \sqrt{r_{\rm H} M/2}$, we have:
\begin{equation}
    t_{\rm in} \sim \frac{2\sqrt{2}}{3} \frac{(r_{\rm H}/M)^{3/2}}{\alpha (h/r_{\rm H})^2}  \Big(1-\sqrt{\frac{r_{\rm ISCO}}{r_{\rm H}}} \Big) M.
\end{equation}

For an equal-mass binary, setting $\alpha=0.1$, $(h/r_{\rm H}) =0.1$, and $r_{\rm H} \sim 0.3 a$ we obtain $t_{\rm in} \sim 150 \times (a/M)^{3/2} (1-\sqrt{r_{\rm ISCO}/r_{\rm H}}) M$, which is a rough upper limit to the inflow time.
This crude estimate, which assumes that accretion occurs due to viscous stresses, must be compared with the timescales associated with CBD accretion.
The largest influx of mass from the CBD is related to the motion of the lump, which has a Keplerian frequency at the inner edge.
Accretion from the lump to the black holes occurs then with a timescale given by $t_{\rm acc} = 5 P_{\rm bin} = 5 \times 2 \pi/\Omega_{\rm bin} = 10 \pi (a/M)^{3/2} M$. 
We see here that for $a < 20\,M$, the timescales start to be comparable, $t_{\rm in}/t_{\rm acc} \sim 1$, in agreement with 3D GRMHD simulations \citep{combi2022minidisk}.

Starting from very small separations, $a = 10\,M$, Refs. \cite{Farris:2011PhRvD, Gold_2014b} showed using full numerical relativity GRMHD simulations that mini-disks do not form in the system; instead, accretion occurs directly via streams from the inner edge of the CBD to the black holes. These results hold for different equations of states and with/without cooling. We should notice, however, that the accretion picture might change if decoupling occurs before those separations are reached. 

For slightly larger initial separations, $a = 20\,M$, Bowen et al. \citep{Bowen17} first found using a cooled, relaxed circumbinary accretion disk (taking as initial data the final state of the simulation in Ref. \cite{Noble12}), that mini-disks are in fact present, but their mass and accretion rate are highly variable. This is due to the inflow time being comparable to the accretion timescales, as we showed above, and as discussed in Ref. \cite{Gold_2014a}. When the mini-disk enters this starvation mode (or a filling and depleting cycle), the specific angular momentum of the gas falling into the black holes determines whether the material will form a mini-disk or a transient structure. 

In general, the gas falling into the cavity from the CBD needs to have low specific angular momentum $\ell_{\rm in}$ to pass the centrifugal barrier of the binary. Refs. \cite{Shi2015} and \cite{Tiede2021} analyzed the distribution of specific angular momentum of the gas that falls into the cavity using Newtonian 3D MHD and 2D VH respectively. In particular, Tiede et al. \cite{Tiede2021} showed that the specific angular momentum distribution of the gas that is accreted by the binary is $\ell_{\rm in} \sim 2-0.5\, \sqrt{a/M} M $. Once the infalling matter is captured by one of the black holes, the fate of the stream is decided by whether $l_{\rm in}$ is high enough to sustain orbital motion around the black hole. We thus need to compare $l_{\rm in}$, transformed to the frame of the black hole, with the specific angular momentum at the ISCO, $l_{\rm ISCO} \sim 2\sqrt{3} (M/2)$. 

The angular momentum budget of mini-disk for small separation binaries was investigated using GRMHD simulations and an approximated strong-field metric in Ref. \cite{combi2022minidisk} for spinning and non-spinning black holes, and in Ref. \cite{avara2023accretion} for non-spinning black holes. These two papers found that mini-disks at these separations have two states: a disk-dominated phase, where the material orbits a number of times the black hole, and a stream-dominated phase, where most of the mass is advected directly to the black hole. For spinning black holes, the mini-disk is larger, which allows for more thermal radiation due to turbulent dissipation \citep{Gutierrez_etal2022}. Moreover, Avara et al. \citep{avara2023accretion} showed that at these separations hydrodynamic stresses dominate the angular momentum transport.

Paschalidis et al. \cite{paschalidis2021minidisk} were the first to demonstrate, using equal-mass 3D GRMHD simulations of hot flows, that mini-disk mass increases with higher spin black holes.


\section{Magnetic fields and outflows}
\label{sec:outflows}

Magnetic fields are a crucial ingredient in the engine of an AGN.
On a (Keplerian) rotationally supported flow, magnetic fields can trigger the development of turbulent stresses through the MRI driving accretion, and they are also responsible for launching powerful outflows that take energy and angular momentum out from the system. The rotating spacetime of a single black hole can interact with the large-scale fields present in the flow \citep{liska2020large,beckwith2009transport} and produce relativistic jets through the Blandford--Znajeck (BZ) mechanism \citep{blanford1977}. Many GRMHD simulations have been performed in the last 20 years showing how jets are launched and how they interact with the surrounding medium \citep{davis2020magnetohydrodynamics}.

The BZ mechanism is an example of how curved spacetime can drastically affect the dynamics of the magnetic field \citep{gralla2014spacetime}. In a binary black hole spacetime, the acceleration of the black holes, the quadrupole gravitational field, and the spin of the black holes can all contribute to generating a Poynting flux that propagates energy to infinity. This has been demonstrated in the context of force-free plasma analytically \citep{yang2016plasma} and in various numerical relativistic simulations \citep{palenzuela2010dual, alic2012accurate, moesta2012detectability}. These simulations evolved a binary on a uniform force-free plasma with a vertical magnetic field, showing that a double jet structure emerges during inspiral, and the Poynting flux increases as they approach the merger. The EM luminosity in this binary merger scenario has two contributions due to spin (BZ) and orbital motion, which is a strong function of velocity :
\begin{equation}
    L_{\rm EM} \sim L_{\rm bin} + L_{\rm spin}, \quad L_{\rm bin} \propto v^2 \sim a^2 \Omega_{\rm K}^2, \quad L_{\rm spin} \sim a_\bullet^2,
\end{equation}
where $a_\bullet$ is the black hole spin.
This also happens in single-boosted black holes, which acquire an effective ergosphere and generate a Poynting flux as well \citep{neilsen2011boosting,cayuso2019astrophysical,penna2015energy}. Ref. \cite{moesta2012detectability} found that the Poynting flux produced by the binary can be divided into a collimated and uncollimated part, which arises from the spin and quadrupole field respectively, see the analysis done in Ref. \cite{yang2016plasma}. 

A similar scenario was explored in MHD simulations for a uniform plasma with an initial $\beta \sim P/b^2 \sim 10^2$ \citep{giacomazzo2012, Kelly2017} including work with different spin inclinations \citep{cattorini2021fully, cattorini2022misaligned}. These simulations show a similar phenomenology to force-free: a magnetically-dominated funnel producing a Poytning flux above and below the black hole spin axis, increasing during the merger. Even though a magnetized region is formed, these simulations do not seem to produce outflows but rather inflows, most likely because they start at very small separations, $a \sim 10M$ and there is no time to produce a proper jet before the merger. We notice that simulations of single black holes embedded in uniform magnetized gas have indeed shown jet production, e.g.,  \cite{ressler2021magnetically}.

\begin{figure}
    \centering
    \includegraphics[width=.6\linewidth]{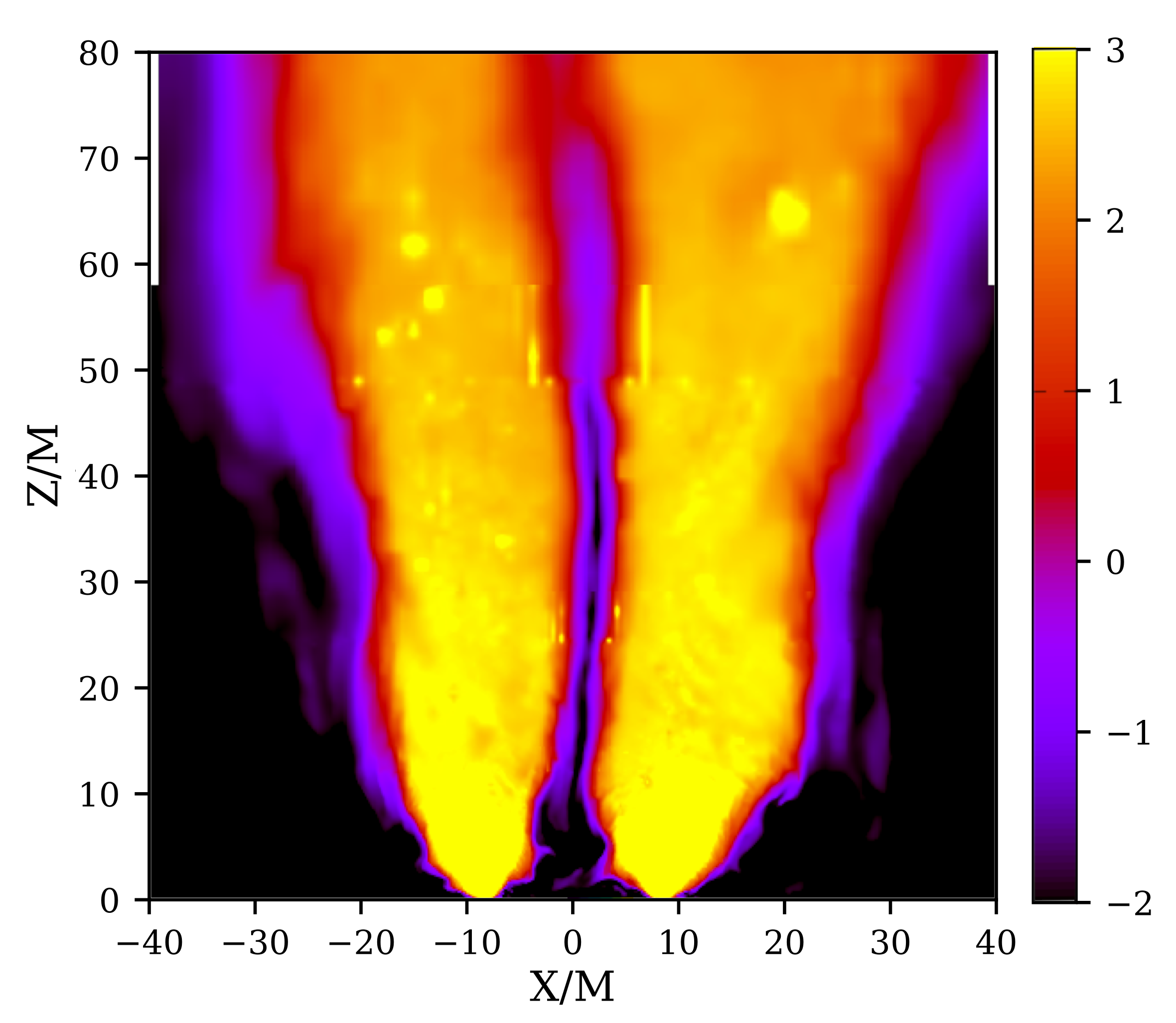}
         \caption{Magnetization ($\sigma = b^2/2\rho )$ for orbiting black holes in a full 3D GRMHD simulation of equal-mass binaries with aligned spines of $a=0.7$. Figure from Ref. \cite{bright2023minidisc}.}
         \label{fig:jets}
\end{figure}

The non-linear interaction between magnetic fields and a binary spacetime in a force-free environment ($b^2/\rho \gg 1$) will only occur if the cavity is strongly depleted and the disk still provides sufficient magnetic flux, e.g. at late stages of decoupling (although it is still unclear whether the density can drop to reach a force-free state).
On the other hand, the mildly magnetized ($b^2/\rho \sim 1$) uniform gas-cloud scenario could well be realized in underfed AGNs or the interior regions of super-Eddington radiatively inefficient flows; however little radiation will emerge from the accretion in these flows and thus will be difficult to detect in distant galaxies.

Unlike the uniform plasma scenario, most luminous AGNs with a binary engine may have an accretion structure described in earlier sections: a cooled CBD that feeds the black holes and forms mini-disks. To understand what sort of outflows these systems produce, we should ask how large-scale magnetic fields are generated and advected onto the binary. 

In this scenario, the magnetic field must be accreted from the CBD. Contrary to single black hole disks where viscous stresses drive the flow near the black hole ISCO, the gas falls into the cavity through thin streams, which compress and amplify the field. Because the lump orbiting the edge of the CBD is a low-magnetized region, the field topology remains an open question. In this direction, Avara et al. \cite{avara2023accretion} showed that the magnetic flux seems to change polarity at the black hole for non-spinning black holes. Noble et al. \cite{noble2021} showed that greater magnetization weakens the lump; this has been shown using an excised cavity (i.e. simulating only the outer disk) and manually injecting flux. 

Three-dimensional GRMHD simulations of spinning binary black holes starting with a CBD have been performed in Refs. \cite{combi2022minidisk, paschalidis2021minidisk}.
Paschalidis et al. \cite{paschalidis2021minidisk} evolved a hot accretion torus starting with an initial magnetic seed loop in the CBD for $\sim 20$ orbits; see Figure \ref{fig:jets}.
For spinning black holes with $a_\bullet= \pm 0.7$, aligned and anti-aligned, they show the emergence of double Poynting-dominated jets, with efficiencies of $\eta \sim 10 \%$ as in single black hole jets; they find that the maximum jet efficiency is reached for the aligned case, as expected. Combi et al. \cite{combi2022minidisk} evolved a cooled, quasi-steady-state CBD starting also with poloidal magnetic field seeds.
In this work, for spinning black holes with $a=0.6$, it was also shown that the efficiencies are of the same order $\eta \sim 10 \%$.
The same work demonstrated that the Poynting luminosity acquires the same variability as the accretion rate due to the filling and depleting state in the mini-disks, which might translate to an observable in jet emission.

In Ref. \cite{bright2023minidisc}, it was remarked and demonstrated that the strength of the variability in the Poynting flux depends on the spin of the black holes: higher spins lead to more steady disks and steady luminosity.
This was expected from previous arguments since the variability depends on the accretion periodicity of the CBD and the inflow time \citep{Bowen17}, which in turn, depends on the size of the mini-disks, and thus the radius of the ISCO.
To reduce the variability of the accretion rate and the Poynting flux, the inflow time must be much larger than the accretion timescales, see Section 1.5.6. One must be careful here because the accretion timescale onto the cavity is very sensitive to the initial data and thermodynamics of the disk. For instance, the larger accretion timescale is associated with the lump $t_{\rm acc} \sim 5 P_{\rm bin}$ in a cooled accretion disk; a non-cooled accretion disk however does not show this variability. More simulations with realistic (steady-state) initial conditions are needed to address these questions. In addition, for short orbital separations, \cite{avara2023accretion} showed that the magnetic flux is less variable than the accretion rate in the mini-disk.

\section{Electromagnetic signatures and variability}
\label{sec:em}

As we have discussed throughout this chapter, MBHBs surrounded by accretion flows are promising EM emitters.
Nevertheless, large uncertainties still exist on the properties of the EM emission and signatures may serve to distinguish them from conventional AGNs with single black holes.
In what follows, we summarize what numerical simulations have taught us about the EM emission from MBHBs and how this depends on the binary and accretion disk properties.
For more extensive reviews on the EM search of MBHBs, both from a theoretical and an observational point of view, the reader may see Refs. \cite{Bogdanovic_etal2022, DOrazioCharisi2023, BurkeSpolaor_etal2018, WangLi2020, AmaroSeoane_etal2023}

Several techniques are used to search for accreting MBHBs during the various stages of their evolution.
At earlier phases in the merger of the hosting galaxies, the pair of black holes are at large separations and they can appear in the sky as two independent AGNs but in the same galactic nucleus.
The best candidate for such a {\it dual AGN} is found in the radio galaxy 0402+379 \citep{Rodriguez2006, Rodriguez_etal2009, Morganti_etal2009}, but tens of these sources have been identified so far (see Ref. \cite{Bogdanovic_etal2022} and references therein).

MBHBs at later stages of the evolution are much more difficult to distinguish from single AGNs, and we must rely on periodic modulations in the emission or specific spectral features.
Currently, though there are several candidates (e.g., see \citep{DOrazio15, Valtonen_etal2008, Graham2015a, Hu_etal2020, ONeill2022}), none of them have been unambiguously confirmed as a close MBHB (see Ref. \cite{DOrazioCharisi2023}, Sec. 5, for an up-to-date list of candidates and the mechanism by which they were detected).
Theoretical studies predict that a fraction $<10^{-2}$ of AGNs at low redshift ($z < 0.6$) may host MBHBs \citep{Kelley2019} (this fraction increases at higher redshifts \citep{Volonteri_etal2022}).
However, only a fraction of these binaries could be effectively identified as such through its EM emission \citep{Volonteri_etal2009, Krolik2019, DongPaez_etal2023}.

The main distinctive characteristic we expect for the EM emission from close MBHBs is a periodic modulation related to the orbital motion of the SMBHs.
This can be caused both by {\it i)} {\it relativistic effects} such as Doppler {\it beaming} of the emission due to the motion of the mini-disks attached to the black holes \citep{Graham2015b, DOrazio15, charisi2018testing} or self-lensing flares for edge-on systems where one of the black holes passes in front of the other and largely amplifies its emission \citep{Dorazio2018, Hu_etal2020, kelly2021electromagnetic, Ingram_etal2021, DavelaarHaiman2022a, DavelaarHaiman2022b, Porter_etal2024}; or by {\it ii)} {\it intrinsic variability} in the emitting plasma (accretion disk or jets) \citep{Farris:2015MNRAS, dAscoli2018, Tang2018, WesternacherSchneider_etal2022, Gutierrez_etal2022}.

Finally, when the MBHB merges due to the emission of GWs \citep{Campanelli:2006fg}, other characteristic EM signals may appear, although these predictions are much less robust than those listed above.
Possibilities include disruption of emission from the jet \citep{Liu2003}, immediate thermal radiation limited by Eddington \citep{Krolik_2010}, and a variety of signals due to black hole recoil resulting from the merger \citep{Campanelli:2007apj, ONeill09, Rossi09, volonteri2008off, blecha2016recoiling}.
See Ref. \cite{Bogdanovic_etal2022} for a more extensive discussion on EM signatures at the time of the merger.

\subsection{Quasi-periodicities in the emission}

Numerical simulations have shown that the accretion rate onto an MBHB can be strongly modulated, see Sec. \ref{sec:minidisks} and Figure \ref{fig:mdot_ditt}.
This is a natural consequence of the influence that the binary motion (which is periodic) exerts on the accretion flow.
The amplitude and period of these modulations may depend on the binary properties, such as mass-ratio and eccentricity or black hole spin, and intrinsic physical properties of the accretion disk, such as the magnetization or the cooling.

D'Orazio et al. \cite{DOrazio13, DOrazio16} performed 2D hydrodynamical simulations to investigate the influence of the mass ratio on the EM modulation.
The main difference is mediated by the presence (or not) of a lump in the CBD, and the influence it exerts on how fast matter flows towards the cavity and the mini-disks.
The lump typically forms for $q \gtrsim 0.2$.
Since the cavity is predicted to have some degree of eccentricity, every time the lump approaches the pericenter of its orbit ($P\sim 5P_{\rm bin}$), it overfeeds the binary, which in turn increases the accretion rate and the luminosity.
This has been confirmed in numerous 2D VH studies and it has also been found in GRMHD simulations \citep{Noble:2009ApJ, lopezarmengol2021, combi2022minidisk}.
On the other hand, when the CBD lacks a lump ($q \lesssim 0.2$), this long-period modulation disappears, though modulations at the orbital period and its harmonics persist.
These are caused by the interaction of the periodic thrown-back of matter towards the CBD through the streams.

Another property that influences the variability is the eccentricity of the orbit, but for the same fact: for orbits with high eccentricity the lump does not form and the modulations are similar to the circular $q<0.2$ case, namely at the orbital period and their harmonics.
Nevertheless, the variabilities become more bursty for very high eccentricities \citep{Zrake2021, Siwek_etal2023}.
Ref. \cite{westernacher2021} predict modulations on the orbital period that are in phase with the orbit.

Other properties that may alter how emission from accretion MBHB varies are the misalignment between the CBD and binary angular momenta
\citep{Dogan_etal2015, Moody2019, Smallwood_etal2022} or variable Fe K$\alpha$ lines due to relativistic effects \citep{YuLu2001, McKernan_etal2013, Jovanovic_etal2014}.

Finally, jet emission may dominate over that of the accretion disk during some stages of the evolution or at specific wavelengths.
The precession of the jets due to the influence of the binary has been proposed as an explanation of sinusoidal modulation observed in some sources \citep{ONeill2022}.
Finally, for close binaries with mass ratios $q \approx 1$, Guti\'errez et al. \cite{Gutierrez_etal2024} proposed that the two jets may interact between them giving rise to flares of nonthermal emission.

\subsection{Spectral signatures}

Another possibility for the identification of MBHB is by looking at differences in their spectral energy distribution (SED) with respect to standard AGNs, though this may be much more difficult to observe.
One predicted signature is the presence of a ``notch'' in the thermal spectrum \citep{GultekinMiller2012, Roedig:2014}: 
the truncation of the CBD and the presence of a low-density cavity would predict a lack of thermal emission at high energies; if minidisks are present in the cavity and are bright, their emission could cover the highest energy part of the spectrum, leaving a ``notch'' between the thermal peaks of the CBD and the minidisks.
This prediction relies on the assumption that the minidisks and the CBD radiate efficiently, while the {\it streams} that link both components do not.
If these conditions are not fulfilled, the spectrum may lack a clear notch \citep{Gutierrez_etal2022}.

The presence of shocks both in the CBD and in the mini-disks may change the disk temperature profile and produce a thermal spectrum peaking at higher energies, typically in X-rays \citep{Roedig:2014}.
Viscous hydrodynamical simulations with optically thick cooling do predict a high-energy excess but this may be due to the assumption of optically thick emission in the cavity, which may not be true \citep{Farris:2015MNRAS, Tang2018, WesternacherSchneider_etal2022}.

\begin{figure}
    \centering
    \includegraphics[width=0.47\textwidth]{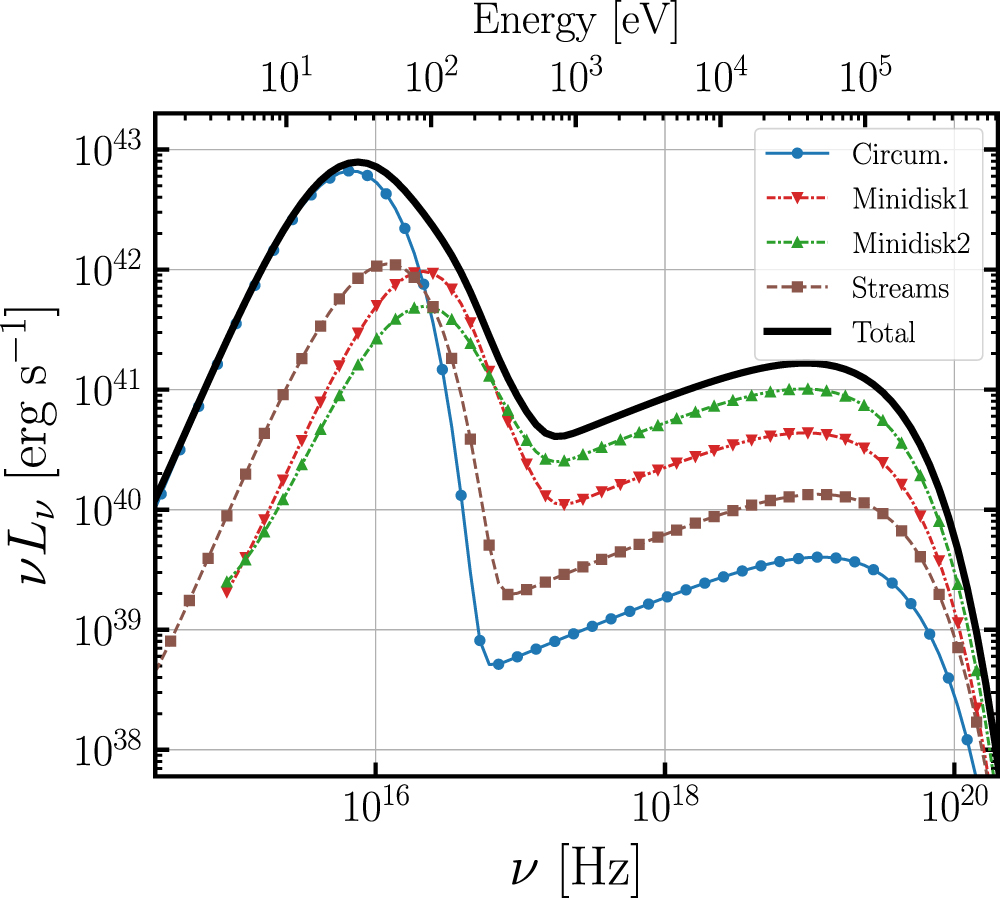}
    \includegraphics[width=0.50\textwidth]{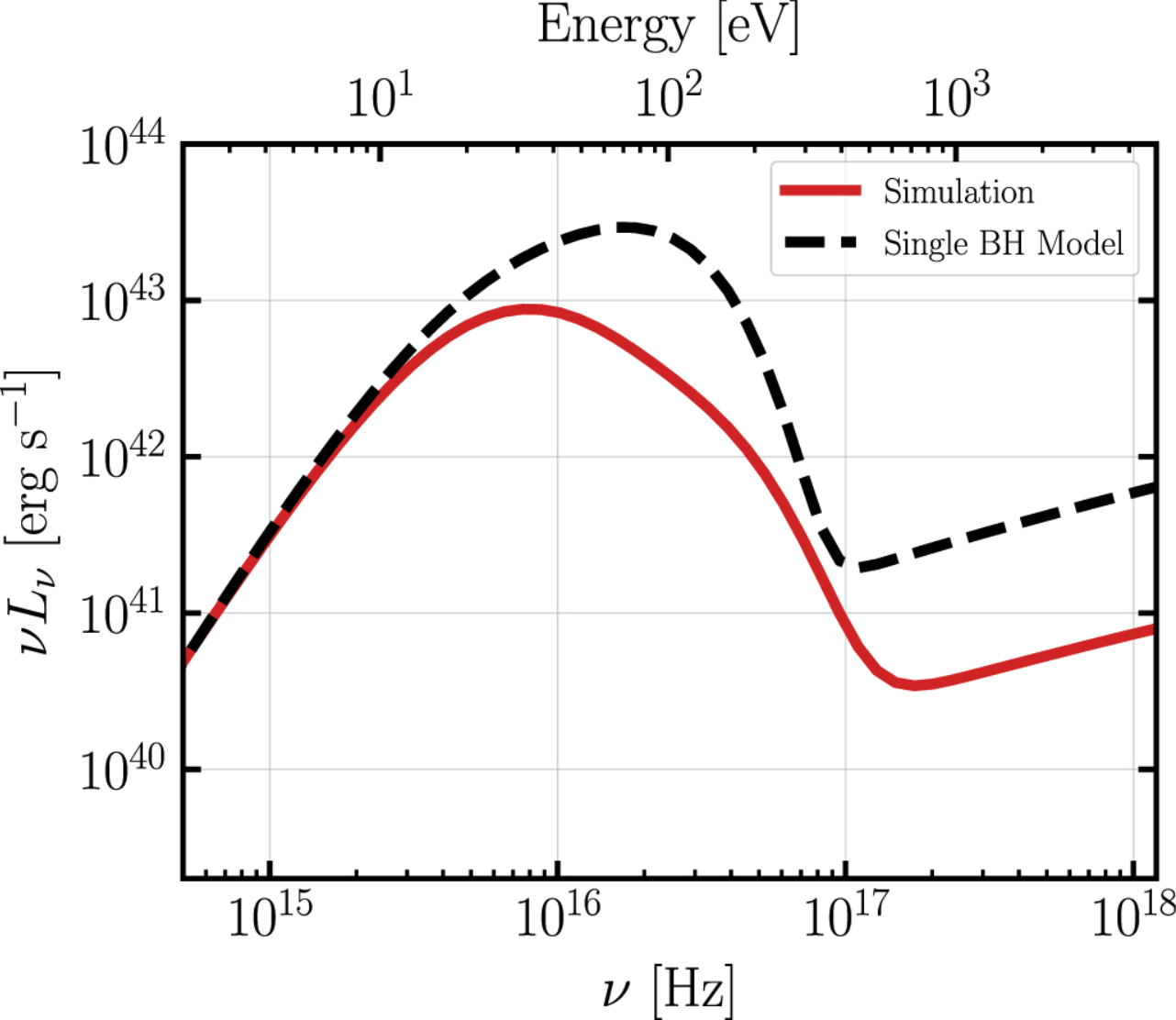}
    \caption{SED from the accretion flow around a binary with equal-mass black holes with spin $a=0.6$.
    {\it Left panel:} Average spectrum for the different components contributing to the
spectrum.
    {\it Right panel:} Total SED of the MBHB compared with an NT model around a single black hole for the same accretion rate. Figure from Ref. \cite{Gutierrez_etal2022}.}
    \label{fig:SED}
\end{figure}

Most simulations link the luminosity of the disk to the accretion rate.
Though this is the case for a standard steady-state disk, it is not clear that the observed luminosity in accreting MBHBs will directly track variations in the simulation-measured accretion.
If the accretion timescale is longer than the modulation period for the accretion rate, then the mini-disks may buffer the accretion rate variability and decrease the amplitude of the variability.
GRMHD simulations of mini-disk accretion \citep{bright2023minidisc, combi2022minidisk} have shown that higher spins produce more massive and larger minidisks, and the viscous timescale becomes longer than the re-filling period.
Though these mini-disks are brighter than for non-spinning black holes, the variability is damped.
The damping depends, however, on the accretion timescales and so it is not clear at what point the variability disappears. Shocks can also induce periodic modulations which depend on the thermodynamics of the gas.

d'Ascoli et al. \cite{dAscoli2018} used the ray tracing code {\tt Bothros} to predict spectra and light curves of GRMHD simulations performed with {\tt Harm3D} \citep{Noble:2009ApJ}.
Guti\'errez et al. \cite{Gutierrez_etal2022} extended this approach and studied in detail the spectral and timing properties in the luminosity of accretion flows onto close binaries for both spinning and nonspinning black holes \citep{Bowen17,Bowen2019,combi2022minidisk}.
In these simulations, the cooling function derived from the GRMHD simulations is transformed into a two-component spectrum composed of a blackbody component, associated to the optically-thick emission from the disk, and a power-law component, which represents the inverse Compton emission usually observed in AGNs.
The left panel of Figure \ref{fig:SED} shows the averaged SED from the simulation with spinning black holes in Ref. \cite{Gutierrez_etal2022}.
The CBD dominates the emission at lower frequencies whereas the minidisks dominate at higher energies.
The emission from the streams falls in
the middle of these two regimes and is comparable in brightness to that of the mini-disks.
The right panel shows the total SED (CBD+streams+mini-disks) compared to the thermal spectrum from an ``equivalent'' (same mass and accretion rate) standard Novikov--Thorne (NT) disk \citep{NovikovThorne73} around a single black hole.
In the binary case, the SED peaks at a
lower frequency and the spectrum at frequencies above the peak is a broken power law instead of a single
decaying exponential.
The overall luminosity is lower in the binary scenario mainly due to the smaller radiative efficiency of mini-disks with respect to standard NT disks.
This is because part of the material falling towards the mini-disks has very little angular momentum and thus it can fall into the black hole without suffering very much dissipation.

\section{The future: open issues}
\label{sec:conclusions}

The coalescence of MBHBs is among the most promising events of multimessenger emission for the upcoming years.
Hence, it is essential to build a clear understanding of how accretion develops in these systems and what is their EM emission.
At the moment, it is still very hard to unambiguously identify such systems and differentiate them from single AGNs.

Accretion flows in MBHBs are more complex than those around single black holes, mainly due to the non-axisymmetric and non-stationary nature of the spacetime, which significantly affects the accretion in the innermost regions where most of the EM emission comes out.
This also implies that analytical modeling is largely limited, and we need to rely much more on numerical simulations.
Moreover, simulating accretion onto MBHBs is itself more complicated and/or more expensive than for single black hole accretion; for this reason, a large majority of the simulations carried out have been 2D Newtonian viscous hydrodynamic simulations.
GRMHD simulations have started to be performed in the last years, though with significant limitations in the duration of the simulation or the range of the parameter space that can be covered.
GRMHD simulations are essential to understand the interplay of accretion with magnetic fields, to study the production of jets, and to accurately predict the EM emission from the innermost regions of the disks or during the latest stages of the binary evolution, including the merger and the post-merger.

Despite these limitations, we have built robust knowledge about some of the properties of accretion flows around MBHBs.
We know that the CBD is typically truncated at a distance $r\sim(2-3) a$, giving rise to a cavity that may be eccentric.
Part of the matter that falls from the CBD towards the cavity is thrown back to the CBD and part of it falls towards the black holes.
If the latter has enough angular momentum, it can form mini-disks around the black holes, and if the black holes have large spins, they can launch jets.
The matter that comes back to the CBD shocks its inner edge and produces the formation of an overdensity, called the lump, 
that orbits the cavity at approximately the distance of CBD truncation.

The lump plays a large role in mediating the amount of matter falling toward the cavity for a near-equal mass binary.
For instance, when one of the black holes passes close to the lump, or when this is at the pericenter of its orbit, accretion towards the cavity increases.
These phenomena give rise to a rich set of periodicities in the accretion rate and mass of the mini-disks, which may then translate to the EM emission as a modulated signal.

In the next years, we will need to improve the capabilities of GRMHD simulations to obtain more accurate and realistic predictions and to expand our coverage of the parameter space. For instance, full 3D GRMHD simulations \cite{ressler2024black} using a semi-analytical metric are now starting to investigate inclined binaries and related effects such as spin-orbit coupling and jet precession. Some of the topics that need to be explored more in-depth are

\begin{itemize}
    \item How do the initial conditions for the simulations (disk profile and magnetic field topology and amplitude) influence the disk and jet evolution and the EM outcome?
    \item How long does the disk take to reach a clear steady state? Are there differences with respect to results obtained from hydrodynamic simulations?
    \item How does the global picture of accreting MBHB change if the disk is misaligned with respect to the orbit?
    \item Do MBHBs produce jets? How frequent is jet production compared to a single-black hole scenario? If there are two jets, do they interact?
    \item How does a realistic thermodynamic and radiation transport treatment influence the properties of the CBD and the mini-disks?
    \item Are there unique signatures in the SED of accreting MBHBs?
\end{itemize}

\begin{acknowledgement}
We thank Manuela Campanelli, Julian Krolik, Scott Noble, Huan Yang, Sean Ressler, and Bart Ripperda for fruitful discussions.
EG acknowledges funding from the National Science Foundation under grant No AST-2108467 and from an Institute for Gravitation and the Cosmos fellowship.
This research was enabled in part by support provided by SciNet (www.scinethpc.ca) and Compute Canada (www.computecanada.ca). LC is a CITA National fellow and acknowledges the support of the Natural Sciences and Engineering Research Council of Canada (NSERC), funding reference DIS-2022-568580. 
Perimeter Institute for Theoretical Physics.
Research at Perimeter Institute is supported in part by the Government of Canada through the Department of Innovation, Science and Economic Development Canada and by the Province of Ontario through the Ministry of Colleges and Universities. 

\end{acknowledgement}


\bibliography{bhm_references}
\bibliographystyle{plain}


\end{document}